\documentclass{article}

\usepackage{tabularx}
\usepackage{booktabs}
\usepackage{PRIMEarxiv}
\usepackage{amsmath} 
\usepackage[utf8]{inputenc} 
\usepackage[T1]{fontenc}    
\usepackage{hyperref}       
\usepackage{url}            
\usepackage{booktabs}       
\usepackage{amsfonts}       
\usepackage{nicefrac}       
\usepackage{microtype}      
\usepackage{lipsum}
\usepackage{fancyhdr}       
\usepackage{graphicx}       
\graphicspath{{media/}}     
\usepackage[most]{tcolorbox}
\usepackage[most]{tcolorbox}
\usepackage{xcolor}
\usepackage{multirow}
\usepackage{makecell}
\usepackage{tabularx}
\usepackage{booktabs}

\newcommand{\dist}[5]{%
\makecell[l]{%
\textsc{I}: #1\\
\textsc{L}: #2\\
\textsc{M}: #3\\
\textsc{H}: #4\\
\textsc{C}: #5%
}}

\newtcolorbox[auto counter, number within=section]{examplebox}[2][]{
  colback=gray!3,
  colframe=black!40,
  fonttitle=\bfseries,
  title=Example~\thetcbcounter: #2,
  label={#1},
  breakable,
  enhanced,
  boxrule=0.5pt,
  arc=1.5mm,
  left=1.5mm,
  right=1.5mm,
  top=1.2mm,
  bottom=1.2mm
}

\newcommand{\boxsep}{%
  \par\medskip
  {\color{black!25}\hrule height 0.4pt}
  \medskip
}

\usepackage{titlesec}

\titlespacing*{\section}
{0pt}        
{1.2ex}      
{0.8ex}      

\title{AI Native Asset Intelligence
}

\author{
  \textbf{Gal Engelberg},
  \textbf{Leon Goldberg},
  \textbf{Konstantin Koutsyi},
  \textbf{Boris Plotnikov} \\
  \textbf{Tiltan Gilat,}
  \textbf{Ben Benhemo} \\
    Sola Security, Tel Aviv, Israel \\
  \texttt{Corresponding author: gal.e@sola.security}
}

\begin{document}
\maketitle

\begin{abstract}
Modern security environments generate large volumes of fragmented signals across cloud resources, identities, configurations, and third-party security tools. Although AI-native security assistants improve interactive access to this data, they remain largely reactive: users must ask the right questions, manually inspect candidate assets, and interpret disconnected findings. This interaction model does not scale to enterprise environments, where the significance of a security signal depends on exposure, exploitability, dependency structure, and business context. As a result, repeated queries over the same environment may produce unstable prioritization because the AI system lacks a structured, reusable basis for comparing assets and grounding its reasoning.
This paper introduces \emph{AI-native asset intelligence}, a framework for transforming heterogeneous security data into a structured intelligence layer that supports consistent, contextual, and proactive asset-level reasoning. The first contribution is an AI-native conceptual architecture that combines a modeling layer, which constructs a representation of assets, identities, relationships, controls, attack vectors, and blast-radius patterns, with a scoring layer that converts fragmented signals into a normalized measure of asset importance. This intelligence layer provides AI systems with an explicit reasoning substrate, enabling both reactive exploration and proactive insight generation. The second contribution is a context-aware scoring system that separates intrinsic exposure from contextual importance. Intrinsic exposure captures misconfiguration findings and attack-vector evidence, while contextual importance captures anomaly, blast radius, business-function criticality, and data criticality. AI-based contextualization is used to interpret asset-specific evidence, such as adjusting finding severity and classifying business or data importance, while bounded deterministic aggregation preserves consistency and auditability. These components are combined through bounded, monotone, and saturating functions to produce an interpretable final score in $[0,1]$.
We evaluate the scoring system on a production-environment snapshot containing $131{,}625$ resources across 15 vendors and 178 asset types. The evaluation focuses on behavioral validation through sensitivity analyses and ablations. The results show that the scoring components behave as intended: severity mappings control finding sensitivity, AI-based severity adjustment refines prioritization without broadly creating extreme scores, attack-vector scoring responds to rare exploitability evidence, and contextual modulation selectively modifies exposed resources based on business or data importance. These findings support AI-native asset intelligence as a principled foundation for stable prioritization and proactive security-posture reasoning.
\end{abstract}


\section{Introduction}

Recent advances in AI-native security tools have improved the ability of practitioners to query, analyze, and interpret security data. However, these systems remain fundamentally \textit{reactive}, relying on users to ask the right questions, investigate specific assets, and manually explore potential issues. This interaction model does not scale to modern enterprise environments, where the volume, velocity, and complexity of security-relevant data far exceed what can be explored through manual querying. As a result, there is a growing need for AI-native systems that can move beyond reactive interaction and support continuous, automated understanding of security posture.

Enabling such capabilities is inherently challenging. At its core lies the difficulty of accurately assessing enterprise security posture, particularly in cloud- and identity-centric environments. These environments generate large volumes of signals, primarily misconfigurations and access control issues, that are fragmented across multiple systems and evaluated in isolation \cite{liu2024misconfig, shao2026framework}. However, these signals are typically associated with intrinsic severity scores that fail to capture their true impact. In practice, the significance of a finding is highly context-dependent, shaped by factors such as exposure, privilege relationships, dependency structures, and business criticality. Consequently, isolated findings provide only a partial and often misleading view of security posture, making it difficult to consistently determine which asset is more critical than the other.

This challenge reveals a more fundamental limitation: the inability of AI systems to reliably reason over fragmented security data when answering security posture questions, such as identifying the top risky assets (e.g., “what are the top 5 risky S3 buckets?”). Existing approaches, including AI assistants and code-driven agents, operate directly on heterogeneous and disconnected signals without a structured abstraction that defines how these signals should be aggregated and compared. As a result, reasoning is unstable and non-systematic. Repeated queries over the same environment may yield inconsistent and conflicting results, as the model lacks a principled basis for ranking. This instability reduces trust in AI outputs and constrains systems to reactive usage, as reliable and reproducible inference is a prerequisite for proactive analysis.

We argue that these limitations stem from the absence of an \textit{intelligence layer above raw security data}. By introducing a structured representation of assets and a principled, context-aware scoring mechanism, such a layer enables AI systems to perform systematic, consistent, and holistic reasoning across assets and their relationships. Crucially, once reasoning becomes reliable, proactive behavior naturally emerges: the system can continuously evaluate the environment, identify high-importance assets, and surface insights without requiring explicit user queries. In this sense, proactive AI is not an independent capability, but a direct consequence of structured and consistent inference.

In this work, we introduce \textit{AI-native asset intelligence}, a conceptual framework that establishes an intelligence layer above raw security data to enable systematic, consistent, and holistic reasoning across assets, identities, and their relationships. Central to this framework is a context-aware scoring system that aggregates heterogeneous signals, capturing both intrinsic exposure (e.g., misconfigurations and exploitability) and contextual factors (e.g., blast radius and business criticality), into a unified and interpretable measure of asset criticality. Together, the proposed architecture and scoring mechanism provide a principled foundation for reliable AI reasoning over fragmented security data, enabling consistent ranking and supporting proactive insight generation.

The rest of the paper is organized as follows. Section~\ref{sec:background} presents background and related work. Section~\ref{sec:architecture} introduces the conceptual architecture for AI-native asset intelligence. Section~\ref{sec:scoring} describes the proposed context-aware scoring system. Section~\ref{sec:ablation} presents an experimental study analyzing the behavior and contribution of the system components.

\section{Background}\label{sec:background}

Work most relevant to \emph{AI-native asset intelligence} spans five connected areas: asset criticality and prioritization frameworks, contextual reasoning over assets and identities, scoring and aggregation methods, structured abstractions for machine reasoning, and AI systems for security posture analysis. Across these areas, a consistent gap emerges: while prior work improves individual components of prioritization or reasoning, it does not provide a unified intelligence layer that transforms fragmented security signals into a canonical, reusable representation for asset-level reasoning.

Traditional prioritization has been dominated by vulnerability-centric standards. CVSS remains the most widely used language for vulnerability severity and, in version 4.0, explicitly separates base, threat, environmental, and supplemental metrics \cite{first_cvss40}. EPSS complements this view by estimating the probability that a published CVE will be exploited in the wild, thereby adding a data-driven threat signal to prioritization workflows \cite{jacobs_epss_2021, first_epss_2026}. SSVC reframes prioritization as a stakeholder-specific decision process rather than a universal scalar score \cite{spring_ssvc_2021}. In parallel, NIST criticality and enterprise-risk documents connect prioritization to mission importance, business impact, and enterprise response planning \cite{paulsen_criticality_2018, quinn_8286b_2025, quinn_8286d_2025}. These frameworks establish a clear consensus that severity alone is insufficient, yet they remain incomplete for AI-native reasoning. CVSS and EPSS are vulnerability-centered, SSVC is decision-oriented rather than representation-oriented, and NIST frameworks focus on governance rather than machine-operational abstractions. Critiques such as Howland's analysis of CVSS further highlight the limitations of severity-centric prioritization \cite{howland_cvss_2023}. As a result, existing approaches identify relevant dimensions of context but do not yield a reusable notion of asset importance suitable for AI systems.

A complementary body of work shows that security posture is inherently relational. Graph-based approaches model how exposure propagates through dependencies, permissions, and trust relationships. Elmiger et al.\ use graph analysis to identify critical attack paths in cloud environments \cite{elmiger_graphs_2024}, while GRASP models serverless policies as reachability graphs to reason about exposure paths \cite{polinsky_grasp_2024}. Gouglidis et al.\ apply model checking to verify IAM policies \cite{gouglidis_googleiam_2023}, and Gl\"ockler et al.\ emphasize the structural complexity of enterprise IAM \cite{glockler_iam_2024}. More broadly, graph-mining surveys argue that preserving relationships among cyber entities is essential for accurate reasoning \cite{yan_graphsurvey_2024}. Although these approaches demonstrate the importance of context, they are typically designed for specific tasks such as path discovery or policy verification. They do not provide a unified abstraction that supports consistent asset-level reasoning across heterogeneous security questions.

Scoring and aggregation methods attempt to bridge raw findings and operational prioritization by incorporating contextual signals. CAVP introduces a context-aware vulnerability prioritization model \cite{jung_cavp_2022}, while SmartPatch addresses prioritization in interdependent environments \cite{yadav_smartpatch_2022}. Kure et al.\ model asset criticality directly in cyber-physical systems \cite{kure_assetcriticality_2022}, and recent surveys synthesize prioritization into severity, exploitability, contextual, and predictive dimensions \cite{jiang_vulnprior_2025}. Despite these advances, most approaches remain vulnerability-centric or remediation-centric. Context is often treated as metadata rather than as structured relationships among assets, identities, and dependencies. This limits their applicability to AI systems, which require stable, shared semantics across diverse queries rather than task-specific scoring mechanisms.

The closest precursor to an intelligence layer appears in the literature on cybersecurity knowledge graphs and neuro-symbolic reasoning. Cybersecurity knowledge graphs integrate heterogeneous data into unified representations for situational awareness and reasoning \cite{sikos_cyberkg_2023, zhao_ckg_2024}. Systems such as CyberGraph and CTINexus demonstrate how structured representations can be constructed and enriched using both standardized data and LLM-based extraction \cite{falcarin_cybergraph_2024, cheng_ctinexus_2024}. Parallel work on combining knowledge graphs with LLMs shows that structured representations improve reasoning quality, interpretability, and faithfulness \cite{luo_rog_2024, ji_kgllm_2024, cheng_neuralsymbolic_2024}. However, most of this work focuses either on CTI integration or on general reasoning benchmarks, rather than on operational security posture analysis. Consequently, it does not provide a canonical abstraction for asset-level prioritization and ranking.

Recent work on AI for security posture highlights the growing role of LLMs and agentic systems. LLMs are increasingly applied to vulnerability analysis, malware analysis, intrusion detection, and identity security posture management \cite{xu_llm4security_2025,engelberg2026solaVisibilityISPM}, while agent-based systems such as CORTEX and LLM-PD explore collaborative and proactive defense workflows \cite{wei_cortex_2025, zhou_llmpd_2024}. Practical deployments in SOC environments further demonstrate the potential of agentic AI \cite{banstola_agentic_2026}. At the same time, the literature shows that LLM-based systems remain sensitive to prompts and evaluation conditions. Studies on prompt sensitivity and evaluation variability highlight challenges in achieving stable and reproducible outputs \cite{zhuo_prosa_2024, polo_prompteval_2024, hua_promptartifact_2025}, while work on adversarial robustness underscores the need for defenses against prompt injection \cite{ali_securecai_2026}. These findings suggest that reliable AI-driven posture reasoning requires structured intermediate representations that stabilize inference.

Taken together, prior work establishes that effective prioritization requires combining severity, exploitability, context, and business impact; that security posture is fundamentally relational; and that structured representations improve AI reasoning. However, these insights remain fragmented across separate frameworks and systems. What is missing is a unifying abstraction that connects raw security data to AI reasoning: an intelligence layer that models assets, identities, dependencies, and contextual importance in a canonical form, and enables consistent asset-level ranking across both reactive queries and proactive workflows. This gap motivates the proposed AI-native asset intelligence framework.

\section{Conceptual Architecture}\label{sec:architecture}

\begin{figure*}[ht!]
    \centering
    \includegraphics[width=1.0\textwidth]{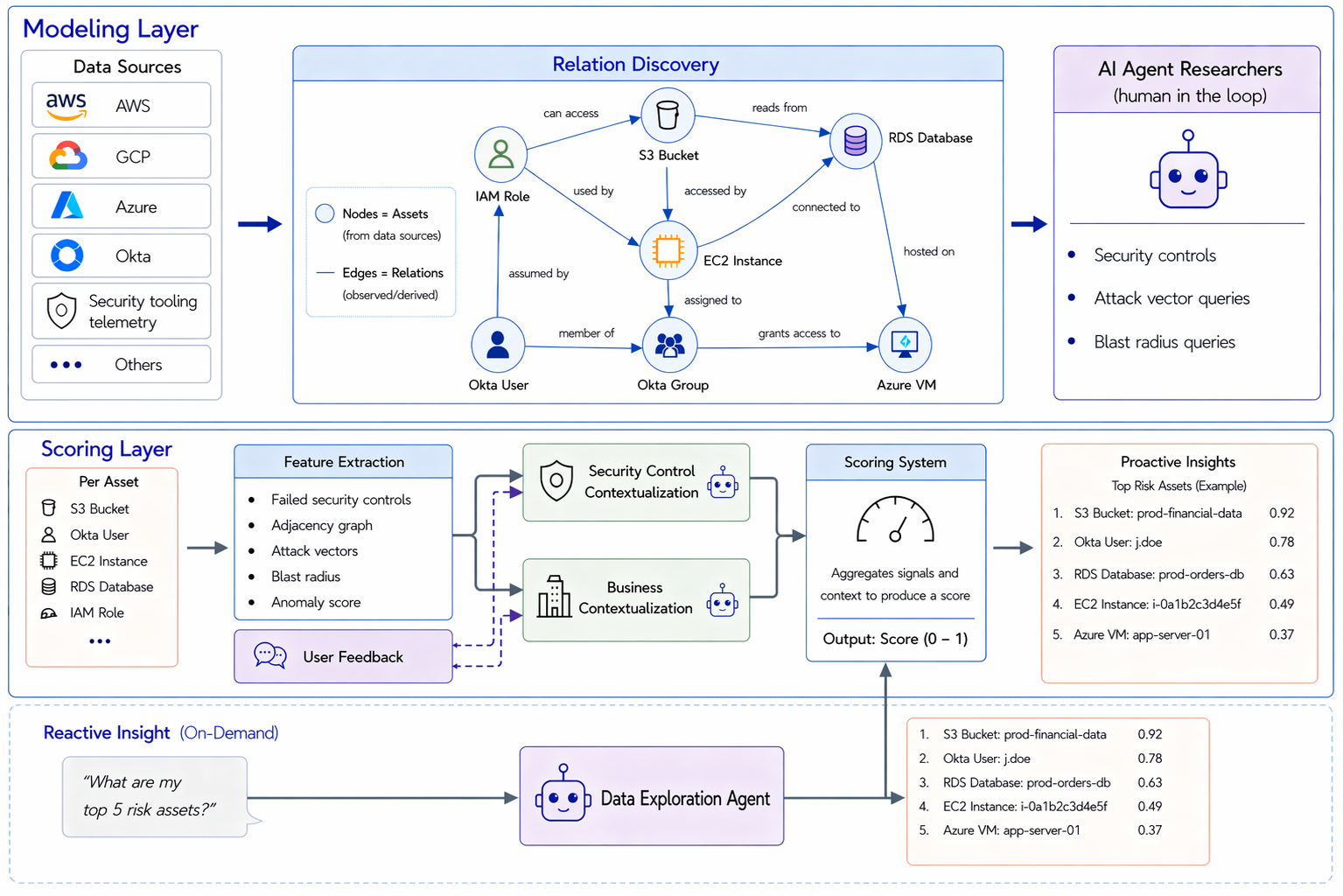}
    \caption{AI-Native Asset Intelligence}
    \label{fig:conceptual_architecture}
\end{figure*}

We organize the system into two cooperating layers: a modeling layer that constructs a unified, semantically rich representation of an organization's assets, and a scoring layer that converts this representation into calibrated, contextualized signals. Both layers feed a dual-mode insights interface that supports proactive surfacing of high-risk assets and reactive, on-demand exploration. Figure \ref{fig:conceptual_architecture} summarizes the flow.

\subsection{Modeling Layer}

\paragraph{Heterogeneous data ingestion.}

The modeling layer ingests configuration, identity, and telemetry data from a heterogeneous set of providers: public-cloud control planes (AWS, GCP, Azure), identity providers (e.g., Okta), code-hosting platforms, and third-party security tooling. Provider-specific extractors normalize raw API responses into a vendor-agnostic tabular form and persist them to an immutable, versioned columnar store, so that every downstream computation is reproducible against an explicit snapshot of the asset.

\paragraph{Relation discovery.}

Tabular records are lifted into a typed property graph whose nodes denote assets (identities, workloads, data stores, network endpoints, code artifacts) and whose edges denote relations between them (ownership, identity assumption, data flow, network reachability, configuration attachment, control attestation). Edges are produced through three complementary mechanisms: (i) deterministic, schema-driven rules that encode well-known intra-vendor relationships and policy semantics; (ii) cross-vendor joins on normalized identifiers (principals, e-mail addresses, repository owners) that bridge otherwise-disconnected silos; and (iii) language-model–assisted edge synthesis, in which candidate relations are proposed from the joint schema and validated against actual data. The output is a single multi-tenant property graph in which any asset can be reached from any other through a chain of typed, semantically meaningful edges, regardless of its originating vendor.

\paragraph{Knowledge overlays via agentic researchers.}

On top of the asset graph and underlying raw data, we maintain three classes of security knowledge artifacts as first-class graph overlays: \emph{security controls}, \emph{attack vectors}, and \emph{blast-radius patterns}. Security controls are declarative predicates that encode whether a given asset satisfies a defined security posture requirement. Attack vectors are parameterized graph-traversal patterns that model an adversary's reachability or privilege-escalation strategy from an initial foothold to a target asset. Blast-radius patterns are graph traversals that, given a compromised asset, enumerate downstream assets whose confidentiality, integrity, or availability may be impacted. Importantly, these artifacts operate at the \emph{metadata level}: they are not materialized results, but executable, queryable specifications that are evaluated at runtime over the underlying data. The intelligence layer therefore maintains the \emph{definitions} of these artifacts, while their execution produces context-specific outputs on demand.

Rather than manually authoring these artifacts, we treat their generation as the responsibility of an agentic AI researcher. This researcher agent operates in a closed loop over three complementary sources of context. First, it samples the underlying tabular data to learn the concrete schemas, value distributions, and resource-specific idioms present in the environment. Second, it retrieves and grounds its hypotheses in public security knowledge, including vendor documentation, best-practice corpora, and standardized threat-intelligence taxonomies such as MITRE ATT\&CK~\cite{mitre_attack}, Common Weakness Enumeration (CWE)~\cite{mitre_cwe}, and Common Attack Pattern Enumeration and Classification (CAPEC)~\cite{mitre_capec}. Third, it interacts directly with the asset graph via structured tool calls (e.g., neighborhood exploration, path finding, and edge-type enumeration) to verify that proposed patterns are realizable within the modeled environment.

Given these inputs, the agent synthesizes candidate artifacts in the form of parameterized graph traversals or tabular predicates, accompanied by natural-language descriptions, taxonomy-aligned tags, and self-assessed confidence scores. Each candidate undergoes multi-stage validation: structural validation ensures well-formedness, bounded traversal depth, and schema consistency; semantic validation requires the agent to justify why the constructed pattern constitutes a valid attack vector, control violation, or blast-radius relationship; and deduplication is performed via embedding-based similarity against previously accepted artifacts. Candidates that satisfy these criteria are promoted into the overlay and become immediately available to downstream components.

This design decouples the platform, data ingestion, graph construction, scoring, and delivery, from the evolving body of security knowledge it operates over, while preserving a strong human-in-the-loop validation process. Security practitioners act as \emph{scenario curators}, guiding the system by proposing relevant threat models, operational questions, and security hypotheses that should be captured as artifacts. At the same time, all generated knowledge artifacts undergo careful manual review and validation by domain experts before promotion into the intelligence layer. This ensures that the resulting controls, attack vectors, and blast-radius queries maintain high fidelity, correctness, and alignment with real-world security practices. By combining automated generation with expert validation, the system enables scalable growth of the knowledge base while maintaining the accuracy, reliability, and auditability required for security-critical applications.

\subsection{Scoring Layer}

The scoring layer is responsible for transforming the structured but heterogeneous representation produced by the modeling layer into a consistent notion of asset importance. This transformation is not purely computational; it is fundamentally a \emph{reasoning problem}. The central challenge is that security posture signals are fragmented, noisy, and context-dependent, and therefore cannot be reliably interpreted in isolation. A high-severity misconfiguration may be irrelevant in one context and critical in another; similarly, assets with identical structural properties may differ significantly in business importance. The proposed conceptual architecture addresses this challenge by explicitly structuring the scoring process as a sequence of reasoning steps, rather than a direct aggregation of signals. The goal is not only to produce a score, but to ensure that the process by which the score is derived is \emph{systematic, consistent, and reproducible}, properties that are essential for both human trust and stable AI inference. Conceptually, the scoring layer follows a three-stage pipeline: \emph{feature extraction}, \emph{AI-native contextualization}, and \emph{aggregation}. Each stage is designed to resolve a specific limitation of naive signal-based scoring.

\paragraph{Feature extraction.}

The modeling and knowledge layers expose a diverse set of signals: failed security controls, attack-vector reachability, blast-radius impact, configuration anomalies, and structural graph features. These signals differ in scale, semantics, and statistical behavior. For example, a control failure carries an ordinal severity, attack vectors accumulate as counts of exploit paths, and blast radius reflects graph reachability. A naive aggregation of these signals would produce inconsistent and unstable results, as no common interpretation exists across them. The feature extraction stage addresses this by transforming each signal into a normalized, unit-interval representation while preserving its semantic meaning. For instance, attack-vector counts are mapped through a saturating function to reflect diminishing returns (the difference between one and three attack paths is significant, while the difference between thirty and forty is not), and blast-radius measures are normalized relative to comparable assets to ensure robustness across environments of different scale. The outcome of this stage is a structured evidence vector per asset, in which heterogeneous signals are made comparable without collapsing their underlying meaning.

\paragraph{AI-native contextualization.} Even after normalization, signals remain under-determined with respect to actual importance. The same finding may correspond to a benign configuration in one environment and a critical exposure in another. This ambiguity is the primary source of both false positives and inconsistent reasoning in existing systems. We address this through \emph{AI-native contextualization}, which explicitly treats interpretation as a reasoning task. Rather than encoding all contextual logic as static rules, we leverage agentic processes that reason over structured and semi-structured evidence, including tabular data, graph structure, and organizational metadata.

Two complementary contextualization processes are applied. First, \emph{security-control contextualization} refines individual findings. For each failing control, an agent evaluates whether the recorded severity reflects the asset’s actual exposure. For example, a public-access configuration on a storage bucket may be appropriate if the bucket is intended to serve public content, but critical if it stores sensitive data. The agent reasons over the asset’s configuration, dependencies, and upstream protections, and may downgrade, retain, or amplify the severity accordingly. This step collapses the long tail of formally valid but operationally irrelevant findings, which would otherwise dominate aggregate scoring.

Second, \emph{business contextualization} evaluates the asset’s importance within the organization. This process classifies the asset along structured dimensions such as functional role, environment, dependency centrality, and data criticality. For example, a database serving as a system of record for customer transactions in a production environment is inherently more important than an identical database used for development testing. Importantly, this classification is not heuristic but grounded in a fixed rubric, ensuring consistency across assets while allowing flexible interpretation of diverse inputs.

A key design principle is that contextualization \emph{does not introduce new signals}; it interprets existing ones. This preserves the integrity of the underlying data while enabling richer reasoning.

\paragraph{Aggregation.}

The final stage combines evidence and contextual signals into a single scalar score. A central design decision is to maintain a strict separation between these two components. Evidence captures directly measurable exposure: misconfigurations, exploit paths, and structural risk. Context captures how important that exposure is, whether it affects a production system, sensitive data, or a central dependency. Treating these components symmetrically would lead to undesirable behavior, such as context overwhelming weak evidence or evidence ignoring critical business implications. Instead, we adopt a multiplicative formulation in which context acts as a bounded modifier of evidence. This ensures that context can amplify or attenuate a score, but cannot override strong or weak evidence entirely. For example, a critical misconfiguration on a low-impact asset remains significant, but does not dominate prioritization; conversely, moderate exposure on a highly critical asset is elevated appropriately.

This design directly addresses the instability observed in AI-based reasoning over raw security data. Without a structured aggregation framework, repeated queries over the same environment may yield inconsistent rankings due to the lack of a principled comparison basis. By enforcing a consistent separation and interaction between evidence and context, the scoring layer produces stable and reproducible outputs.




The scoring layer is therefore not merely a ranking function, but a structured reasoning pipeline. By transforming fragmented signals into comparable evidence, interpreting them through AI-native contextualization, and combining them through a principled aggregation scheme, it enables consistent, interpretable, and reproducible assessment of asset importance. This structure is essential for supporting both proactive insight generation and stable responses to interactive queries, and forms the foundation for the formal quantification model presented in the next section.

\section{Scoring System}\label{sec:scoring}

This section presents the quantification model used to rank assets according to their security importance. The objective is to provide a consistent and interpretable prioritization mechanism across heterogeneous cloud environments, where signals are fragmented across configurations, identities, cloud resources, and security tooling. Security posture is not directly observable. It emerges from multiple interacting factors: local security conditions, such as misconfigurations; exploitability, such as reachable attack vectors; and contextual importance, such as business role, data sensitivity, and blast radius. The scoring model therefore separates \emph{intrinsic exposure} from \emph{contextual importance}, and defines how these two components interact in a controlled and explainable manner.

The model is designed to satisfy four requirements. First, it must be \emph{bounded}, so every score remains in $[0,1]$. Second, it must be \emph{monotone}, so additional evidence of exposure or importance never decreases the score. Third, it must exhibit \emph{diminishing returns}, so many weak signals do not overwhelm a small number of severe ones. Fourth, it must be \emph{context-aware but evidence-preserving}: contextual importance may amplify or attenuate measured exposure, but should not fully replace it.

We model the environment as a set of assets $a \in \mathcal{A}$, where each asset represents an entity that can be misconfigured, compromised, misused, or used as part of an attack. Examples include compute resources, storage buckets, databases, secrets, IAM users, roles, service accounts, cloud projects, subscriptions, and control-plane resources. 
Each asset may be associated with a set of findings. In this work, findings primarily represent security-relevant misconfigurations, such as excessive IAM permissions, public exposure, missing encryption, missing multi-factor authentication, or disabled protective controls. Each finding has an initial severity label drawn from the ordered scale:
\[
\mathcal{S} = \{\textsc{Info}, \textsc{Low}, \textsc{Medium}, \textsc{High}, \textsc{Critical}\}.
\]

\begin{figure*}[ht!]
    \centering    \includegraphics[width=\textwidth]{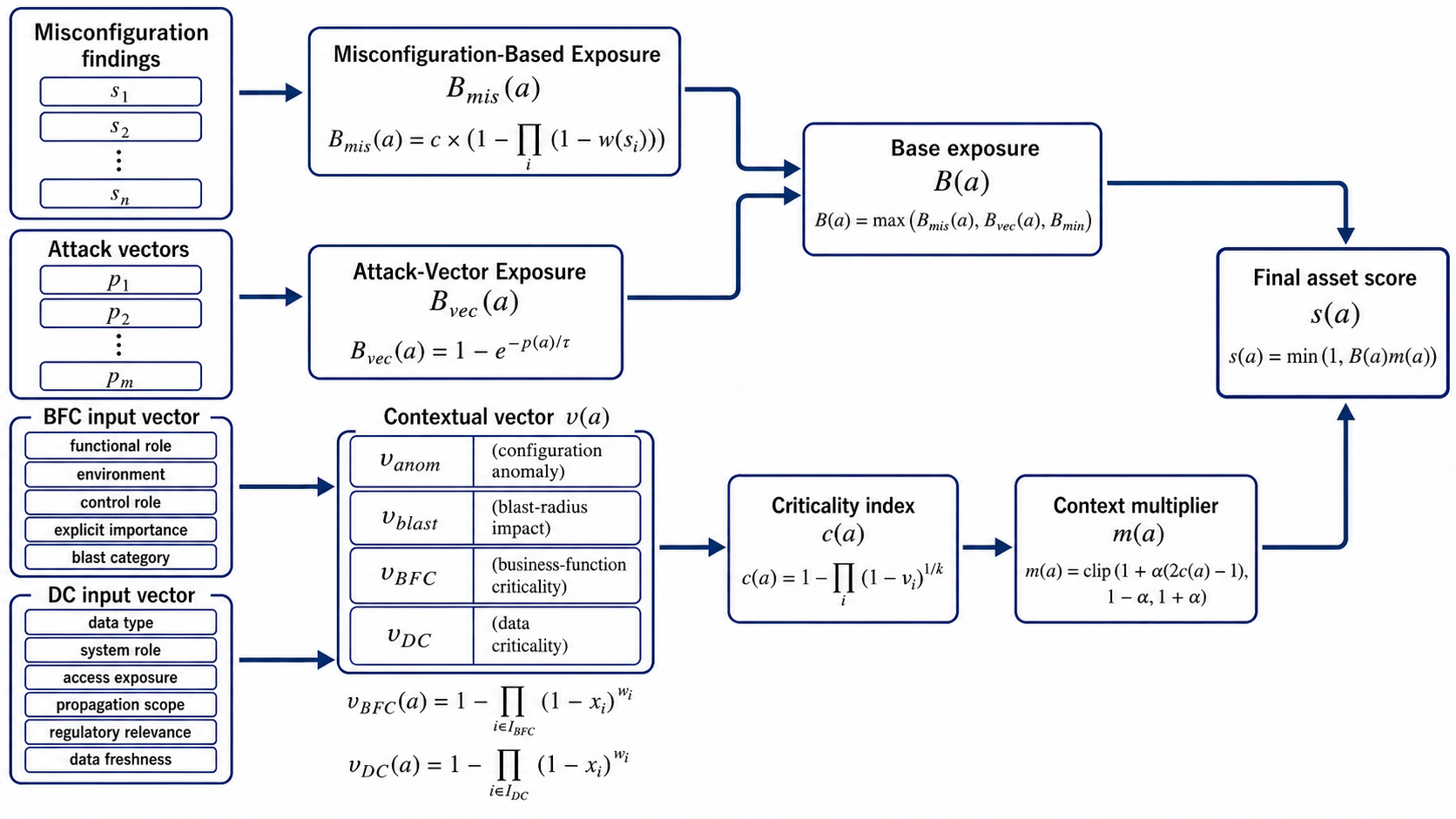}
    \caption{Overview of the asset scoring system}
    \label{fig:scoring_system}
\end{figure*}

In addition to findings, each asset is characterized by structural signals derived from the asset graph. An \emph{attack vector} describes a concrete way in which an adversary may initiate compromise against the asset. A \emph{blast radius} describes the downstream entities that may be affected if the asset is compromised. Finally, contextual signals capture organizational importance, such as whether the asset belongs to production, supports a core business function, or handles regulated data.
Figure~\ref{fig:scoring_system} provides an overview of the complete scoring pipeline, showing how intrinsic exposure and contextual signals are aggregated into the final asset score.

\subsection{Base Exposure}

The base exposure score captures the intrinsic security exposure of an asset before business context is considered. We define it using two independent channels: a misconfiguration channel and an attack-vector channel. The misconfiguration channel captures local weaknesses observed on the asset, while the attack-vector channel captures concrete exploitability patterns in which the asset serves as the subject of compromise.

Let $B_{\text{mis}}(a)$ denote the misconfiguration-based exposure and $B_{\text{vec}}(a)$ denote the attack-vector exposure. The base exposure score is:
\[
B(a) = \max\left(B_{\text{mis}}(a),\; B_{\text{vec}}(a),\; B_{\min}\right).
\]

The use of the maximum is intentional. Misconfigurations and attack vectors are different explanations for why an asset may be security-relevant, but they may also describe overlapping phenomena. Adding them would risk double-counting the same underlying exposure. Taking the maximum ensures that either severe misconfigurations or concrete exploitability can elevate the asset, while preserving a single interpretable base score.

The floor $B_{\min}$ is used only to avoid collapsing sparse assets to zero. This is important in large cloud environments, where the absence of observed findings does not necessarily imply absence of importance or exposure. For example, an IAM role with little direct evidence may still be structurally central in the graph; assigning a minimal nonzero base allows contextual signals to remain meaningful while still keeping the asset below genuinely exposed assets.





\paragraph{Misconfiguration-Based Exposure.}

Let an asset $a$ have misconfiguration findings with severities $\{s_1,\dots,s_n\}$. Each severity is mapped to a weight $w(s_i) \in [0,1]$, where higher severities receive disproportionately larger weights. The misconfiguration exposure is defined as a capped saturating union:
\[
B_{\text{mis}}(a) =
c \cdot \left(1 - \prod_{i=1}^{n} \bigl(1 - w(s_i)\bigr)\right),
\]
where $c \in [0,1]$ is the channel cap.

This form is chosen to satisfy three requirements. It is \emph{bounded} by construction, since the inner product lies in $[0,1]$ and the outer scaling guarantees $B_{\text{mis}}(a) \in [0, c]$. It is \emph{monotone}, since adding a finding with positive weight strictly decreases the inner product, increasing the score. It also exhibits \emph{diminishing returns}, since each additional finding contributes less as the inner product approaches zero and the score approaches the cap.

An additional desirable property follows directly from the multiplicative form: a finding with $w(s_i) = 0$ contributes a factor of $1$ to the inner product, leaving $B_{\text{mis}}(a)$ unchanged. Thus, the score is driven by the magnitude of severity weights rather than by the number of findings alone. Severity dominates locally, while accumulation contributes only through findings with non-negligible weight.

The cap has an important semantic role: \emph{misconfigurations alone should not fully saturate the final asset score}. A long list of local misconfigurations may indicate poor hygiene, but without exploitability or strong contextual importance it should not automatically dominate the ranking. For example, suppose the cap is set near $0.75$. A single critical misconfiguration may push $B_{\text{mis}}(a)$ close to that cap, and the accumulation of several high or medium misconfigurations may also approach it. However, the asset still requires either attack-vector exposure or strong contextual importance to reach the highest priority range. This prevents a low-importance development resource with many local issues from outranking a production identity or data store that is both exploitable and business-critical.

\paragraph{Attack-Vector Exposure.}

Let $p(a)$ denote the number of attack-vector patterns that terminate at asset $a$. The attack-vector exposure is:
\[
B_{\text{vec}}(a) = 1 - e^{-p(a)/\tau},
\]
where $\tau > 0$ controls the saturation rate.

This formulation reflects the fact that attack vectors represent concrete exploitability. The transition from zero to one viable attack path is qualitatively significant: it changes the asset from theoretically misconfigured to practically reachable or exploitable. Additional attack paths increase exposure, but with diminishing effect. For example, an asset with one viable privilege-escalation path should be prioritized substantially higher than an otherwise similar asset with none. However, an asset with twenty paths is not twenty times more important than one with one path; rather, the additional paths increase confidence that the asset is a robust attack target.

Unlike the misconfiguration channel, the attack-vector channel is not capped below one. This is deliberate. Misconfigurations represent local evidence, whereas attack vectors represent realized exploitability in the modeled environment. Therefore, attack-vector exposure may push the base score into the highest range even without additional contextual amplification.

\subsection{Contextualization}

The base score captures exposure, but it does not capture asset importance. In practice, identical misconfigurations may have very different consequences depending on the asset's role. A public-access setting on a development bucket containing test data may be low priority, while the same configuration on a production bucket containing regulated customer data may be urgent. Similarly, a missing MFA control on a low-privilege user is different from the same finding on a cloud administrator or service account with broad access.

To model this distinction, we define a contextual vector:
\[
\mathbf{v}(a) =
[v_{\text{anom}},\; v_{\text{blast}},\; v_{\text{BFC}},\; v_{\text{DC}}],
\]
where $v_{\text{anom}}$ is configuration anomaly, $v_{\text{blast}}$ is normalized blast-radius impact, $v_{\text{BFC}}$ is business-function criticality, and $v_{\text{DC}}$ is data criticality. Each component is normalized to $[0,1]$, where higher values indicate stronger contextual importance. Components that are unavailable for an asset are omitted rather than imputed, so the model aggregates only observed contextual evidence.

\paragraph{Configuration anomaly.}
The configuration anomaly component $v_{\text{anom}}(a)$ measures the extent to which asset $a$ deviates from comparable assets in the same environment. It is a deterministic signal derived from configuration, entitlement, or policy-deviation analysis, and is represented as a continuous value in $[0,1]$. In practice, the value is interpreted as a percentile score: values close to $0$ indicate that the asset is similar to its peer group, while values close to $1$ indicate that the asset is highly unusual relative to comparable assets.

This component captures abnormality rather than direct exposure. An anomalous asset is not necessarily vulnerable, but unusual configurations often indicate exceptional permissions, non-standard access paths, unusual connectivity, or drift from expected organizational baselines. For example, a service account with atypically broad permissions, a storage bucket with uncommon sharing patterns, or a workload with unusual network reachability may receive a high anomaly score. In the contextual vector, anomaly therefore acts as an environmental signal that increases prioritization when technical exposure is also present.

\paragraph{Blast-radius impact.}
The blast-radius component $v_{\text{blast}}(a)$ measures the potential structural scope of impact if asset $a$ is compromised. It is a deterministic graph-derived signal represented as a continuous value in $[0,1]$, computed as a normalized or percentile-based measure of affected downstream entities. Values close to $0$ indicate that compromise is expected to remain local, while values close to $1$ indicate that compromise may affect a large portion of the environment.

This component is distinct from attack-vector exposure. Attack vectors describe how an adversary may initially compromise an asset, whereas blast radius describes what may be affected after compromise succeeds. Thus, $v_{\text{blast}}(a)$ captures impact rather than exploitability. Assets with many dependent resources, broad access relationships, or central positions in the asset graph receive higher blast-radius values because their compromise can affect more users, services, identities, resources, or data stores.

\paragraph{Business-function criticality.}
The business-function criticality component $v_{\text{BFC}}(a)$ captures the operational importance of asset $a$ to the organization. Unlike configuration anomaly and blast radius, which are continuous structural signals, $v_{\text{BFC}}(a)$ is constructed from semantic criteria that describe the asset's role in business operations. These criteria include functional role, dependency centrality, environment, control role, explicit business importance, and blast-radius category. Each criterion is represented using ordered labels mapped to numeric scores $x_i \in [0,1]$, as shown in Table~\ref{tab:bfc_features}. Higher scores indicate stronger business criticality.

The functional-role criterion distinguishes auxiliary assets from internal-support assets, customer-facing support assets, and core-business-function assets. The environment criterion distinguishes development, pre-production, and production assets. Control role captures whether the asset operates as a single workload, shared service, or orchestrator. Explicit business importance captures signals from names, tags, ownership metadata, or other organizational annotations. Finally, blast-radius category captures the semantic class of affected systems, such as internal operations, shared platforms, customer data platforms, customer front doors, systems of record, identity or control-plane resources, and revenue flows.

The purpose of $v_{\text{BFC}}(a)$ is to encode the fact that two assets with the same exposure may differ substantially in operational priority. A production identity-control-plane asset, revenue-flow component, or system of record should be ranked differently from a development helper resource, even if both exhibit similar local findings. High business-function criticality therefore indicates that compromise of the asset is likely to disrupt important organizational capabilities.

\begin{table}[ht!]
\centering
\small
\renewcommand{\arraystretch}{1.15}
\begin{tabularx}{\linewidth}{p{2.7cm}p{5.0cm}X}
\toprule
\textbf{Feature} & \textbf{Ordered labels and scores} & \textbf{Explanation} \\
\midrule

Functional role
& auxiliary $(x_i=0.2)$; internal support $(x_i=0.5)$; customer-facing support $(x_i=0.8)$; core business function $(x_i=1.0)$
& Measures how directly the asset supports business operations, from indirect support to core business functionality. \\
\cmidrule(lr){1-3}

Environment
& development $(x_i=0.3)$; pre-production $(x_i=0.6)$; production $(x_i=1.0)$
& Captures operational stage, with production assets receiving the highest criticality. \\
\cmidrule(lr){1-3}

Control role
& single workload $(x_i=0.4)$; shared service $(x_i=0.7)$; orchestrator $(x_i=1.0)$
& Measures whether the asset affects only itself, supports multiple consumers, or controls other resources. \\
\cmidrule(lr){1-3}

Explicit business importance
& none $(x_i=0.3)$; implicit $(x_i=0.7)$; explicit $(x_i=1.0)$
& Captures business-importance signals from names, tags, ownership, or explicit annotations. \\
\cmidrule(lr){1-3}

Blast-radius category
& internal operations $(x_i=0.4)$; shared platform $(x_i=0.6)$; customer data platform $(x_i=0.7)$; customer front door $(x_i=0.8)$; system of record $(x_i=0.9)$; identity/control plane $(x_i=1.0)$; revenue flow $(x_i=1.0)$
& Captures the most critical semantic category affected through the asset's blast radius. \\

\bottomrule
\end{tabularx}
\caption{Business-function criticality features, ordered labels, and numeric scores.}
\label{tab:bfc_features}
\end{table}

Let $\mathcal{B}_a$ denote the set of business-function criteria available for asset $a$. For each criterion $i \in \mathcal{B}_a$, let $x_i$ be the numeric score assigned according to Table~\ref{tab:bfc_features}. The score is computed using a weighted soft-maximum form:
\[
v_{\text{BFC}}(a)
=
1 - \prod_{i \in \mathcal{B}_a} (1 - x_i)^{\tilde{w}_i},
\qquad
\sum_{i \in \mathcal{B}_a} \tilde{w}_i = 1.
\]

The normalized aggregation weights $\tilde{w}_i$ are obtained from default criterion weights after excluding unavailable or low-confidence criteria:
\[
\tilde{w}_i =
\frac{w_i}{\sum_{j \in \mathcal{B}_a} w_j}.
\]
This aggregation treats each criterion as an independent driver of business importance. A single highly critical signal, such as production environment or identity-control-plane blast category, can substantially increase the score, while multiple moderate signals can reinforce one another without requiring all criteria to be high.

\paragraph{Data criticality.}
The data-criticality component $v_{\text{DC}}(a)$ captures the sensitivity and operational importance of data associated with asset $a$. Unlike configuration anomaly and blast radius, which are continuous structural signals, $v_{\text{DC}}(a)$ is constructed from semantic criteria that describe the nature of the data handled by the asset, its role in the data lifecycle, and the potential consequences of compromise. These criteria include data type, system role, access exposure, propagation scope, regulatory relevance, and data freshness. Each criterion is represented using ordered labels mapped to numeric scores $x_i \in [0,1]$, as shown in Table~\ref{tab:dc_features}. Higher scores indicate higher data criticality.

The data-type criterion distinguishes between assets with no relevant data context, unknown data sensitivity, low-sensitivity data, internal data, business-sensitive data, and regulated or sensitive data. The system-role criterion captures whether the asset acts as a transient processor, derived copy, authoritative source, or system of record. Access exposure captures whether access to the asset is restricted, moderate, or broad. Propagation scope captures whether compromise of the associated data remains local, spans multiple services, or crosses organizational domains. Regulatory relevance captures legal or compliance implications. Finally, data freshness captures whether the data is archival, periodically used, or live operational data.

The purpose of $v_{\text{DC}}(a)$ is to encode the fact that two assets with the same technical exposure may differ substantially in consequence depending on the data they handle. For example, a regulated production database, a live customer-data platform, or an authoritative system of record should be prioritized differently from a transient processor or archival dataset, even if both exhibit similar local findings. High data criticality therefore indicates that compromise of the asset may expose sensitive, regulated, broadly accessible, authoritative, or operationally important data.

\begin{table}[ht!]
\centering
\small
\renewcommand{\arraystretch}{1.15}
\begin{tabularx}{\linewidth}{p{2.7cm}p{5.0cm}X}
\toprule
\textbf{Feature} & \textbf{Ordered labels and scores} & \textbf{Explanation} \\
\midrule

Data type
& not applicable (excluded); unknown $(x_i=0.0)$; low $(x_i=0.25)$; internal $(x_i=0.5)$; business-sensitive $(x_i=0.75)$; regulated/sensitive $(x_i=1.0)$
& Captures the sensitivity of data handled or stored by the asset. \\
\cmidrule(lr){1-3}

System role
& not applicable (excluded); transient $(x_i=0.2)$; derived copy $(x_i=0.5)$; authoritative source $(x_i=0.8)$; system of record $(x_i=1.0)$
& Captures the role of the asset in the data lifecycle, from temporary processing to authoritative storage. \\
\cmidrule(lr){1-3}

Access exposure
& restricted $(x_i=0.3)$; moderate $(x_i=0.6)$; broad $(x_i=1.0)$
& Measures how broadly the asset and its data are accessible to principals or services. \\
\cmidrule(lr){1-3}

Propagation scope
& not applicable (excluded); local $(x_i=0.4)$; multi-service $(x_i=0.7)$; cross-domain $(x_i=1.0)$
& Measures how widely compromise of the associated data may propagate. \\
\cmidrule(lr){1-3}

Regulatory relevance
& not applicable (excluded); none $(x_i=0.2)$; moderate $(x_i=0.6)$; high $(x_i=1.0)$
& Captures legal or compliance implications associated with the data. \\
\cmidrule(lr){1-3}

Data freshness
& not applicable (excluded); archival $(x_i=0.3)$; periodic $(x_i=0.6)$; live operational $(x_i=1.0)$
& Captures the operational importance of the data based on recency and active use. \\

\bottomrule
\end{tabularx}
\caption{Data-criticality features, ordered labels, and numeric scores.}
\label{tab:dc_features}
\end{table}

Let $\mathcal{D}_a$ denote the set of data-criticality criteria applicable to asset $a$. For each criterion $i \in \mathcal{D}_a$, let $x_i$ be the numeric score assigned according to Table~\ref{tab:dc_features}. The score is computed using a weighted soft-maximum form:
\[
v_{\text{DC}}(a)
=
1 - \prod_{i \in \mathcal{D}_a} (1 - x_i)^{\tilde{w}_i},
\qquad
\sum_{i \in \mathcal{D}_a} \tilde{w}_i = 1.
\]

Criteria marked as \emph{not applicable} are excluded from the aggregation rather than treated as zero. This distinction is important because many assets, such as identities, network entities, or compute resources, may not directly store or process data, and should not be penalized for lacking data-specific attributes. In contrast, an \emph{unknown} label is retained with low value when there is evidence that the asset handles data but insufficient evidence to classify its sensitivity. The normalized aggregation weights $\tilde{w}_i$ are obtained from the default criterion weights after excluding unavailable or non-applicable criteria:
\[
\tilde{w}_i =
\frac{w_i}{\sum_{j \in \mathcal{D}_a} w_j}.
\]
This aggregation treats each criterion as an independent driver of data importance. A single highly critical signal, such as regulated data type or system-of-record role, can substantially increase the score, while multiple moderate signals, such as internal data with broad access and multi-service propagation, can reinforce one another without requiring all criteria to be high.

\subsection{Context Modulation}

After computing the contextual vector, we aggregate its available components into a scalar criticality index:
\[
c(a) =
1 - \prod_{i=1}^{k} (1 - v_i)^{1/k},
\]
where $k \leq 4$ is the number of contextual components available for asset $a$.

Equal weighting at this level reflects the fact that the four contextual components are heterogeneous and not directly commensurable. Configuration anomaly, blast radius, business function, and data criticality each capture a different dimension of importance. Rather than impose a global ordering among them, the model lets any sufficiently strong dimension elevate the context index, while still allowing multiple moderate dimensions to reinforce one another.

The criticality index is mapped to a bounded multiplier:
\[
m(a) =
\mathrm{clip}
\left(
1 + \alpha(2c(a)-1),\;
1-\alpha,\;
1+\alpha
\right).
\]

This multiplier enforces the requirement that context modulates exposure rather than replaces it. When $c(a)=0.5$, the multiplier is neutral and the base score is unchanged. When $c(a)>0.5$, context amplifies the base score; when $c(a)<0.5$, context attenuates it. The hyperparameter $\alpha$ bounds the maximum effect of context.

This design produces the intended behavior for mixed cases. A misconfigured but low-importance development asset remains bounded by the misconfiguration channel and may be attenuated by context. A moderately exposed production database that is a system of record and handles regulated data is amplified. An asset with both strong attack-vector exposure and high business importance reaches the top of the ranking. Thus, the highest-priority assets are those where measured exposure and contextual importance agree.

\subsection{Final Score}

The final asset score is:
\[
s(a) = \min(1, B(a)m(a)).
\]

This formulation completes the separation between intrinsic exposure and contextual importance introduced earlier. The base score $B(a)$ determines whether an asset is exposed, while the multiplier $m(a)$ determines how strongly that exposure should be prioritized within the environment.

The multiplicative interaction enforces an \emph{evidence-preserving} property. Contextual importance alone cannot generate risk: if $B(a)$ is close to zero, the final score remains low regardless of $m(a)$. Conversely, when exposure is present, contextual signals act as a scaling factor that differentiates assets with similar technical conditions but different operational significance. This ensures that the ranking reflects both \emph{how exposed} an asset is and \emph{how much it matters} if that exposure is realized.

The use of a bounded multiplier, together with the outer $\min(1,\cdot)$ operator, guarantees that the final score remains within $[0,1]$ while preserving monotonicity. Any increase in exposure or contextual importance can only increase (or leave unchanged) the final score. At the same time, the bounded range prevents extreme contextual values from overwhelming the intrinsic signal, maintaining comparability across assets.

An important consequence of this design is that high scores arise only when exposure and context are aligned. Assets with strong exposure but low contextual importance remain below the highest priority range, while assets with high contextual importance but weak exposure are not artificially elevated. The top of the ranking is therefore reserved for assets that are both meaningfully exposed and operationally critical.

Overall, the final score acts as a normalized, interpretable proxy for asset-level risk, suitable for prioritization across heterogeneous cloud environments. It preserves the structure of the underlying signals while providing a single scalar value that is consistent, bounded, and operationally meaningful.

The complete scoring process is illustrated through the end-to-end running example in Example~\ref{box:running_example}.

\begin{examplebox}[box:running_example]{End-to-end running example: publicly writable S3 bucket with Lambda trigger}

Consider an asset $a^\star$ representing an AWS S3 bucket that allows public access and permits external principals to perform \texttt{PutObject}. The bucket is configured to trigger an AWS Lambda function when new objects are created. Thus, external object uploads may initiate downstream compute execution.

\boxsep

\textbf{Parameters.}
For illustration, we use:
\[
c=0.75,\quad B_{\min}=0.05,\quad \tau=0.6,\quad \alpha=0.3,
\]
with the severity weights:
\[
w(\textsc{High})=0.45,
\qquad
w(\textsc{Critical})=0.75.
\]

\boxsep

\textbf{Base exposure.}
The asset has two findings: public access labeled \textsc{High} and public \texttt{PutObject} labeled \textsc{Critical}. The misconfiguration exposure is:
\[
B_{\text{mis}}(a^\star)
=
0.75 \cdot \bigl(1 - (1-0.45)(1-0.75)\bigr)
=
0.75 \cdot (1 - 0.1375)
\approx 0.647.
\]
The asset also participates in one attack-vector pattern: adversary-controlled object upload into a bucket that triggers downstream Lambda execution. Therefore:
\[
B_{\text{vec}}(a^\star)
=
1-e^{-1/0.6}
\approx 0.811.
\]
The resulting base exposure is:
\[
B(a^\star)
=
\max(0.647,\;0.811,\;0.05)
=
0.811.
\]

\boxsep

\textbf{Contextualization.}
Assume the contextual components obtained from the criteria in Tables~\ref{tab:bfc_features} and~\ref{tab:dc_features} are:
\[
\mathbf{v}(a^\star)
=
[0.80,\;0.70,\;0.74,\;0.68],
\]
where the components correspond to configuration anomaly, blast-radius impact, business-function criticality, and data criticality, respectively.

\boxsep

\textbf{Context modulation.}
Using the contextual vector above:
\[
c(a^\star)
=
1 -
(1-0.80)^{1/4}
(1-0.70)^{1/4}
(1-0.74)^{1/4}
(1-0.68)^{1/4}
\approx 0.73.
\]
The corresponding multiplier is:
\[
m(a^\star)
=
1+0.3(2\cdot0.73-1)
\approx 1.14.
\]

\boxsep

\textbf{Final score.}
The final asset score is:
\[
s(a^\star)
=
\min(1,\;0.811\cdot1.14)
=
0.925.
\]

The score is high because the asset is both intrinsically exposed and contextually important: public object upload provides a concrete attack path through the misconfiguration channel, while the Lambda trigger extends the potential impact beyond the bucket via the attack-vector channel. In this example, $B_{\text{vec}}$ dominates $B_{\text{mis}}$, so the base exposure is set by the attack-vector evidence rather than by the misconfigurations alone.

\end{examplebox}

\section{Experimental Evaluation}\label{sec:ablation}

We evaluate the proposed scoring system through a sequence of ablations and sensitivity studies designed to (i) calibrate each scalar parameter, (ii) isolate the contribution of each component to the final score.
All experiments are conducted on a single production-environment snapshot of a real organization's inventory. The snapshot inventories $131{,}625$ distinct resources and the security signals attached to each: failed security controls, attack-vector path counts, blast-radius metrics, business-context attributes, and an anomaly score. Each resource is observed once, so the data are cross-sectional rather than longitudinal.

The environment is multi-cloud and multi-SaaS, drawn from a single organization and integrating data across $15$ vendors. AWS dominates the inventory at $78.6\%$ of all resources; the remaining $21.4\%$ is spread across business-application SaaS (Salesforce, Jira, HiBob, NetSuite), additional cloud providers (Azure, Azure~AD, GCP), source-control and code-review platforms (GitHub), identity providers (Okta, JumpCloud, Google Workspace), and endpoint/security tooling (Jamf, CrowdStrike, Cloudflare). The full vendor breakdown is reported in Table~\ref{tab:dataset:vendors}.
The snapshot covers $178$ distinct asset types, spanning compute, container imagery, identity primitives, data stores, network edge, source code, and SaaS business objects. Aggregating these asset types into seven logical domains yields the distribution in Table~\ref{tab:dataset:domains}: roughly half the inventory is compute and container artefacts, a quarter is identity and access primitives, and the remaining quarter is split between storage/data, network, source-control, and SaaS objects.

The scoring system operates on resources that carry at least one risk signal. Of the $131{,}625$ resources in the snapshot, $39{,}296$ ($29.9\%$) have at least one failed security control and $8$ have at least one positive attack-vector path count, yielding $39{,}304$ resources in scope.

We organize the evaluation around the main stages of the scoring pipeline. The experiments progressively move from local components to the final score: first examining how misconfiguration evidence is transformed into $B_{\text{mis}}(a)$, then how attack-vector counts affect $B_{\text{vec}}(a)$, how both channels interact in the base exposure score $B(a)$, and finally how contextual modulation changes the transition from base score ($B(a)$) to the final score ($s(a)$). This structure allows the analysis to separate component-level behavior from end-to-end scoring behavior. In addition, we assess the effect of AI-based severity adjustment as a cross-cutting analysis.

\begin{table}[h]
\centering\small
\begin{tabular}{@{}lrrrrr@{}}
\toprule
\textbf{Vendor} & \textbf{\# resources} & \textbf{share} & \textbf{\# asset types} & \textbf{\# in scope} & \textbf{coverage} \\
\midrule
AWS              & 103{,}474 & 78.61\% & 76 & 30{,}149 & 29.14\% \\
Salesforce       &  13{,}491 & 10.25\% &  5 &      56 &  0.42\% \\
Azure            &   6{,}354 &  4.83\% & 30 &     321 &  5.05\% \\
Azure~AD         &   2{,}707 &  2.06\% &  8 &  1{,}773 & 65.50\% \\
GitHub           &   2{,}349 &  1.78\% &  9 &      85 &  3.62\% \\
Cloudflare       &   1{,}198 &  0.91\% &  4 &       0 &  0.00\% \\
Okta             &       613 &  0.47\% & 11 &     382 & 62.32\% \\
GCP              &       524 &  0.40\% & 14 &     522 & 99.62\% \\
JumpCloud        &       400 &  0.30\% &  3 &     280 & 70.00\% \\
Jira~Cloud       &       207 &  0.16\% &  6 &       0 &  0.00\% \\
Google Workspace &        92 &  0.07\% &  7 &      12 & 13.04\% \\
Jamf             &        82 &  0.06\% &  1 &      82 & 100.00\% \\
HiBob            &        68 &  0.05\% &  1 &      68 & 100.00\% \\
CrowdStrike      &        57 &  0.04\% &  2 &      57 & 100.00\% \\
NetSuite         &         9 &  0.01\% &  1 &       0 &  0.00\% \\
\midrule
\textbf{Total}   & \textbf{131{,}625} & \textbf{100.00\%} & \textbf{178} & \textbf{39{,}304} & \textbf{29.86\%} \\
\bottomrule
\end{tabular}
\caption{Vendor breakdown of the evaluation dataset.}
\label{tab:dataset:vendors}
\end{table}

\begin{table}[h]
\centering\small
\begin{tabular}{@{}lrr@{}}
\toprule
\textbf{Logical domain} & \textbf{\# resources} & \textbf{share} \\
\midrule
Compute \& Containers     & 62{,}450 & 47.45\% \\
Identity \& Access        & 36{,}525 & 27.75\% \\
Storage \& Data           & 23{,}272 & 17.68\% \\
Network \& Edge           &  4{,}587 &  3.48\% \\
Source Control \& CI      &  2{,}424 &  1.84\% \\
SaaS / Business Apps      &  1{,}591 &  1.21\% \\
Other                     &     775 &  0.59\% \\
\midrule
\textbf{Total}            & \textbf{131{,}625} & \textbf{100.00\%} \\
\bottomrule
\end{tabular}
\caption{Aggregation of the $178$ asset types into seven logical domains.}
\label{tab:dataset:domains}
\end{table}

\subsection{Sensitivity of the Misconfiguration Channel}
\label{sec:eval_ablation_misconfiguration}

The first analysis evaluates the behavior of the misconfiguration-based exposure score $B_{\text{mis}}(a)$ under different severity-to-score mappings. This analysis isolates the finding channel from the rest of the scoring pipeline: attack-vector exposure, contextual modulation, and the final multiplier are excluded, and only the mapping from finding severities to numeric weights is varied. The purpose is to examine whether the capped saturating aggregation behaves as intended when assets accumulate multiple failed controls of different severities.

We evaluate six severity-weight configurations, each defining a mapping from finding severity to numeric score. For compactness, each configuration is written as a tuple ordered by severity: (\textsc{Info},\textsc{Low},\textsc{Medium},\textsc{High},\textsc{Critical}).

The evaluated configurations are: \emph{baseline} $(0.002,0.02,0.08,0.35,0.70)$, which provides the reference non-linear mapping with a strong gap between low/medium and high/critical severities; \emph{conservative} $(0.0005,0.008,0.035,0.25,0.50)$, which reduces the contribution of all severities and especially of critical findings; \emph{very-conservative} $(0.0002,0.004,0.02,0.18,0.40)$, which further lowers the impact of every severity tier; \emph{ultra-conservative} $(0.0001,0.002,0.012,0.12,0.30)$, which strongly suppresses informational, low, and medium findings and substantially tempers high and critical findings; \emph{aggressive} $(0.002,0.02,0.07,0.45,0.80)$, which increases the contribution of medium, high, and critical findings so that a single severe finding produces a large jump in the score; and \emph{linear} $(0.005,0.05,0.15,0.40,0.70)$, which uses a broader near-linear severity scale where informational and low findings are no longer negligible.

For each severity-weight configuration, we compute $B_{\text{mis}}(a)$ as a function of the number of failed controls associated with an asset cluster. The analysis groups assets according to the \textbf{highest severity} present among their findings and reports how the finding score evolves as the number of failed controls increases. This allows us to observe three behavioral properties of the misconfiguration channel: whether higher-severity findings dominate lower-severity ones, whether additional failed controls increase the score monotonically, and whether the score approaches the cap gradually rather than growing without bound.

\begin{figure}[ht!]
    \centering
    \includegraphics[width=\linewidth]{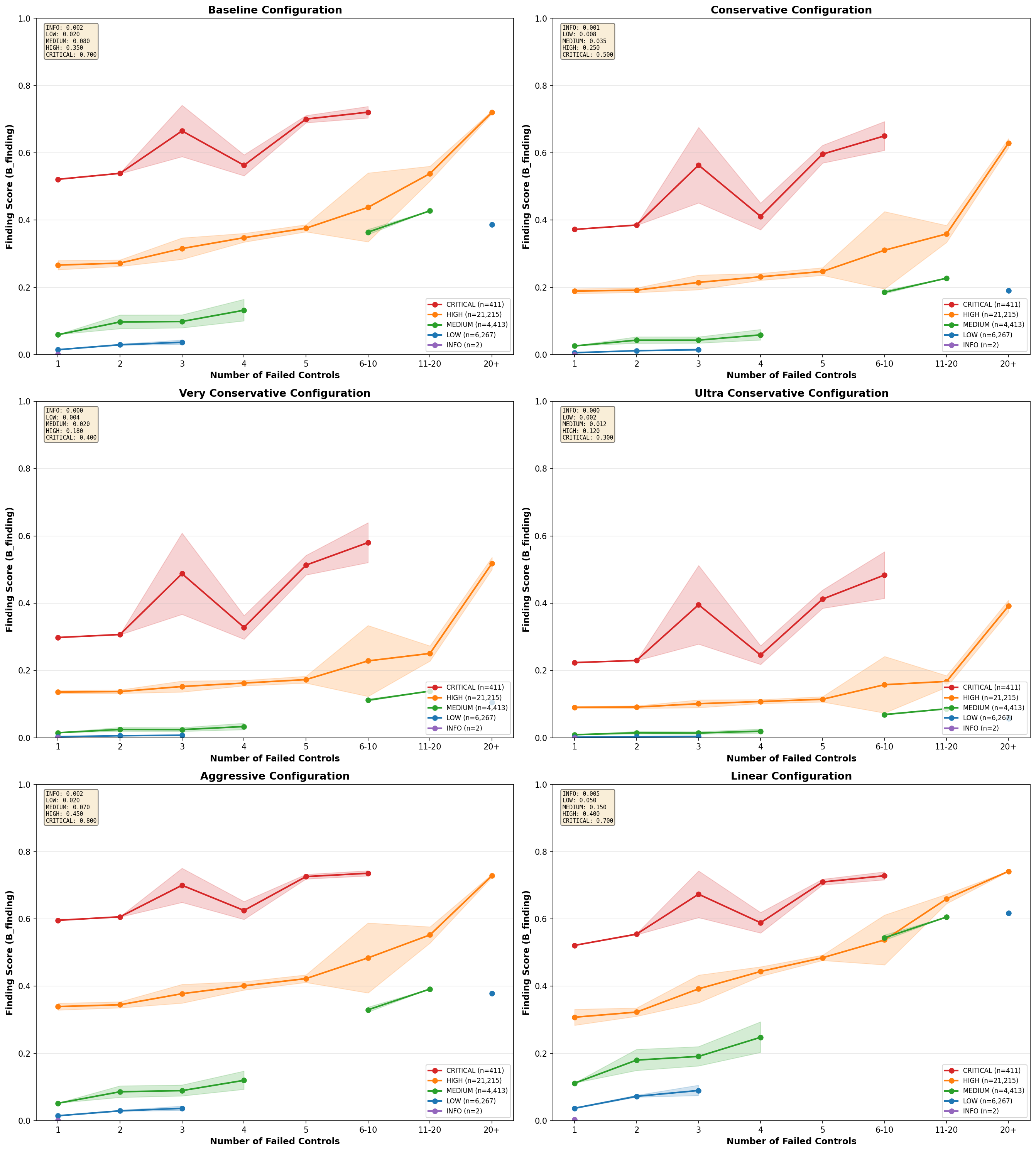}
    \caption{Sensitivity of the misconfiguration exposure score $B_{\text{mis}}(a)$ to the severity-weight configuration.}
    \label{fig:abl1_finding_sensitivity}
\end{figure}

Figure~\ref{fig:abl1_finding_sensitivity} shows the resulting curves for the six configurations. Across all configurations, the score increases as additional failed controls are observed, but the marginal contribution of each additional finding decreases as the score approaches the channel cap. This confirms the intended diminishing-return behavior of the capped saturating formulation. A single high- or critical-severity finding already produces a substantially higher score than a low- or medium-severity finding, while repeated lower-severity findings increase the score more gradually. Because each finding contributes a multiplicative factor of $(1 - w(s_i))$ to the inner product, severities with small weights (informational and low) leave the score nearly unchanged, while severities with large weights drive most of the saturation behavior.

The comparison between configurations shows the expected calibration effect. The aggressive configuration shifts the curves upward, making the finding channel more sensitive to individual findings, especially for high and critical findings: a single critical finding already places the asset well above the medium-severity range. The conservative, very-conservative, and ultra-conservative configurations shift the curves downward, especially for single findings and small numbers of failed controls; under the most conservative settings, even repeated medium-severity findings remain in the lower range of the channel. The baseline configuration provides an intermediate behavior: it gives critical findings a strong effect, but still allows repeated medium and high findings to accumulate gradually. As the number of findings grows, the curves tend to converge toward the channel cap only when the underlying weights are non-negligible, so the asymptote is reached primarily through accumulated severity rather than through finding count alone. Thus, the severity-weight mapping controls both the early-stage sensitivity of $B_{\text{mis}}(a)$ and the rate at which the score approaches the cap, while the cap parameter $c$ bounds its absolute value.

A secondary observation concerns clusters whose highest severity is medium. In some non-linear configurations these clusters can, at large finding counts, reach scores comparable to clusters whose highest severity is critical. This does not indicate a violation of the severity ordering. The grouping is based only on the highest severity observed in the cluster, and does not capture the full composition of findings within the group. A cluster in the medium group may contain many medium findings, while a cluster in the critical group may contain a single critical finding accompanied by mostly informational or low findings. Because the aggregation is cumulative and monotone, a large number of medium findings can legitimately raise the score, and the multiplicative form ensures that this accumulation is driven by severity weight rather than by finding count. Under the corrected aggregation, this effect is most visible in the aggressive and linear configurations and is strongly suppressed in the conservative-family configurations.

The linear configuration behaves differently from the non-linear ones. Because it assigns relatively high weights to all severities, including $w(\textsc{Low})=0.05$ and $w(\textsc{Medium})=0.15$, even moderate accumulations of low- or medium-severity findings raise the score noticeably. Critical and high findings still dominate locally, but the channel becomes more sensitive to repeated low-severity evidence, which can produce inflated scores for assets that exhibit poor configuration hygiene without any genuinely severe finding. This sensitivity reduces the discrimination of the channel between truly severe assets and assets with many minor issues, which is undesirable for prioritization.

Overall, this analysis supports the use of a non-linear severity-to-score mapping together with the multiplicative capped aggregation. The baseline and conservative-family configurations preserve the intended behavior most clearly: severe findings dominate early, repeated findings contribute with diminishing returns, and the misconfiguration channel remains controlled. The aggressive configuration may be useful when the system should prioritize severe findings more strongly, while the linear configuration appears too sensitive to accumulation of low- and medium-severity findings for stable prioritization. The analysis also shows that the multiplicative aggregation makes the channel's behavior driven by the severity weights themselves: severities with small weights leave the score essentially unchanged, so the calibration of the weight tuple directly determines both single-finding impact and asymptotic behavior.

\subsection{Impact of AI-Based Severity Adjustment}
\label{sec:eval_ablation_ai_severity}

The second analysis evaluates the effect of AI-based severity adjustment on the misconfiguration channel. While the previous analysis examined how different severity-to-score mappings affect $B_{\text{mis}}(a)$, this analysis evaluates whether changing the severity labels themselves materially changes the resulting score distribution. The analysis is restricted to resources with at least one failed control and at least one AI-based severity adjustment. This focuses the analysis on the part of the assets where the adjustment mechanism is active, rather than diluting its effect across resources whose severities were not modified. All distributions are resource-weighted, so the reported percentages reflect the share of resources assigned to each score bin.

For each severity-weight configuration, we compute the finding score twice for the adjusted resource subset: once using the original control-level severities, and once using the AI-adjusted severities. We then compare the resulting resource-weighted distributions over the same five score bins used throughout the evaluation: \textsc{Info} $(0$--$0.2)$, \textsc{Low} $(0.2$--$0.4)$, \textsc{Medium} $(0.4$--$0.8)$, \textsc{High} $(0.8$--$0.9)$, and \textsc{Critical} $(0.9$--$1.0)$. This comparison shows how resources move between prioritization ranges as a result of AI-based severity refinement, and whether this movement depends on the selected severity-weight configuration.

Table~\ref{tab:ai_severity_adjustment_explicit} reports the bin-level changes introduced by AI-based severity adjustment. The table shows, for each affected bin, the resource-weighted share before adjustment, the share after adjustment, and the signed change. Positive values indicate that a larger share of resources is assigned to the bin after adjustment, while negative values indicate that resources leave the bin.

\begin{table}[ht!]
\centering
\small
\renewcommand{\arraystretch}{1.15}
\begin{tabular}{llccc}
\toprule
\textbf{Configuration} 
& \textbf{Bin}
& \textbf{Original}
& \textbf{AI-adjusted}
& \textbf{$\Delta$} \\
\midrule

\multirow{3}{*}{Baseline}
& \textsc{Info}
& $0.00\%$
& $53.22\%$
& $+53.22\%$ \\
& \textsc{Low}
& $84.03\%$
& $35.05\%$
& $-48.98\%$ \\
& \textsc{Medium}
& $15.97\%$
& $11.73\%$
& $-4.24\%$ \\

\cmidrule(lr){1-5}

\multirow{3}{*}{Conservative}
& \textsc{Info}
& $23.11\%$
& $62.54\%$
& $+39.43\%$ \\
& \textsc{Low}
& $72.08\%$
& $34.06\%$
& $-38.02\%$ \\
& \textsc{Medium}
& $4.81\%$
& $3.39\%$
& $-1.41\%$ \\

\cmidrule(lr){1-5}

\multirow{3}{*}{Very-conservative}
& \textsc{Info}
& $85.30\%$
& $90.74\%$
& $+5.44\%$ \\
& \textsc{Low}
& $10.88\%$
& $7.56\%$
& $-3.32\%$ \\
& \textsc{Medium}
& $3.82\%$
& $1.70\%$
& $-2.12\%$ \\

\cmidrule(lr){1-5}

\multirow{3}{*}{Ultra-conservative}
& \textsc{Info}
& $89.54\%$
& $93.43\%$
& $+3.89\%$ \\
& \textsc{Low}
& $7.49\%$
& $5.65\%$
& $-1.84\%$ \\
& \textsc{Medium}
& $2.97\%$
& $0.92\%$
& $-2.05\%$ \\

\cmidrule(lr){1-5}

\multirow{3}{*}{Aggressive}
& \textsc{Info}
& $0.00\%$
& $54.06\%$
& $+54.06\%$ \\
& \textsc{Low}
& $66.22\%$
& $17.31\%$
& $-48.90\%$ \\
& \textsc{Medium}
& $33.78\%$
& $28.62\%$
& $-5.16\%$ \\

\cmidrule(lr){1-5}

\multirow{3}{*}{Linear}
& \textsc{Info}
& $0.00\%$
& $26.36\%$
& $+26.36\%$ \\
& \textsc{Low}
& $55.55\%$
& $43.53\%$
& $-12.01\%$ \\
& \textsc{Medium}
& $44.45\%$
& $30.11\%$
& $-14.35\%$ \\

\bottomrule
\end{tabular}
\caption{Explicit resource-weighted bin changes introduced by AI-based severity adjustment. The table reports only bins whose resource share changed after adjustment.}
\label{tab:ai_severity_adjustment_explicit}
\end{table}

The main effect of AI-based severity adjustment is a substantial shift from the \textsc{Low} bin toward the \textsc{Info} bin, with a smaller secondary effect on the \textsc{Medium} bin. Under the baseline configuration, a large share of adjusted resources moves out of \textsc{Low} into \textsc{Info}, while the \textsc{Medium} share decreases by approximately $4\%$. This indicates that the adjustment mechanism primarily refines the prioritization of lower-severity findings, downgrading resources whose finding severities the AI considers less impactful than their original assignment.

The effect remains strong but progressively narrower under more conservative configurations. The conservative configuration shows a similar shift from \textsc{Low} to \textsc{Info}, though slightly attenuated, with an additional small decrease in the \textsc{Medium} bin. Under very-conservative and ultra-conservative configurations, most adjusted resources are already concentrated in the \textsc{Info} bin before adjustment, so the additional movement after adjustment is small. In these settings, the channel is already conservative enough that AI-based adjustment provides only marginal refinement.

The aggressive configuration produces the largest absolute movement, with over half of the adjusted resources moving from \textsc{Low} into \textsc{Info}. The shift is sharper than under the baseline because the aggressive mapping creates a wider gap between low-weight and high-weight severities, so downgrading a finding's severity has a larger effect on its numeric weight.

The linear configuration shows a more even spread between the \textsc{Low} and \textsc{Medium} bins both before and after adjustment. The adjustment redistributes resources across all three lower bins rather than concentrating the movement in a single direction, because the linear mapping assigns relatively high weights to low and medium severities.

Overall, the adjustment mechanism primarily refines prioritization in the lower score ranges, with the dominant effect being a transition from \textsc{Low} to \textsc{Info}. The magnitude of this transition depends on the severity-weight configuration: baseline and aggressive show the strongest effect because their original \textsc{Low}/\textsc{Medium} bins are most populated, while conservative-family configurations already place most resources in the \textsc{Info} bin and therefore exhibit smaller shifts. No configuration produces resources in the \textsc{High} or \textsc{Critical} bins after adjustment, confirming that the AI-based severity refinement does not introduce extreme prioritization; instead, it operates as a fine-grained downgrade mechanism applied to resources whose adjusted severities are less severe than their original labels.

\subsection{Sensitivity of the Attack-Vector Channel}
\label{sec:eval_ablation_attack_vectors}

The third analysis evaluates the behavior of the attack-vector exposure channel under different values of the saturation parameter $\tau$. Unlike misconfiguration findings, which may appear broadly across the resource population, attack vectors represent concrete exploitability patterns. They are therefore expected to be relatively rare, but highly informative when present. For this reason, the analysis is restricted to resources with a positive attack-vector signal, allowing us to examine the behavior of the channel only where exploitability evidence exists.

We evaluate five configurations: $\tau=3$, $\tau=5$, the baseline value $\tau=7$, $\tau=10$, and $\tau=15$. Smaller values of $\tau$ make the channel more sensitive to the first few attack paths, while larger values produce a more gradual increase. The purpose of this analysis is to understand how aggressively the model should react when attack-vector evidence is observed.

\begin{figure}[ht!]
    \centering
    \includegraphics[width=\linewidth]{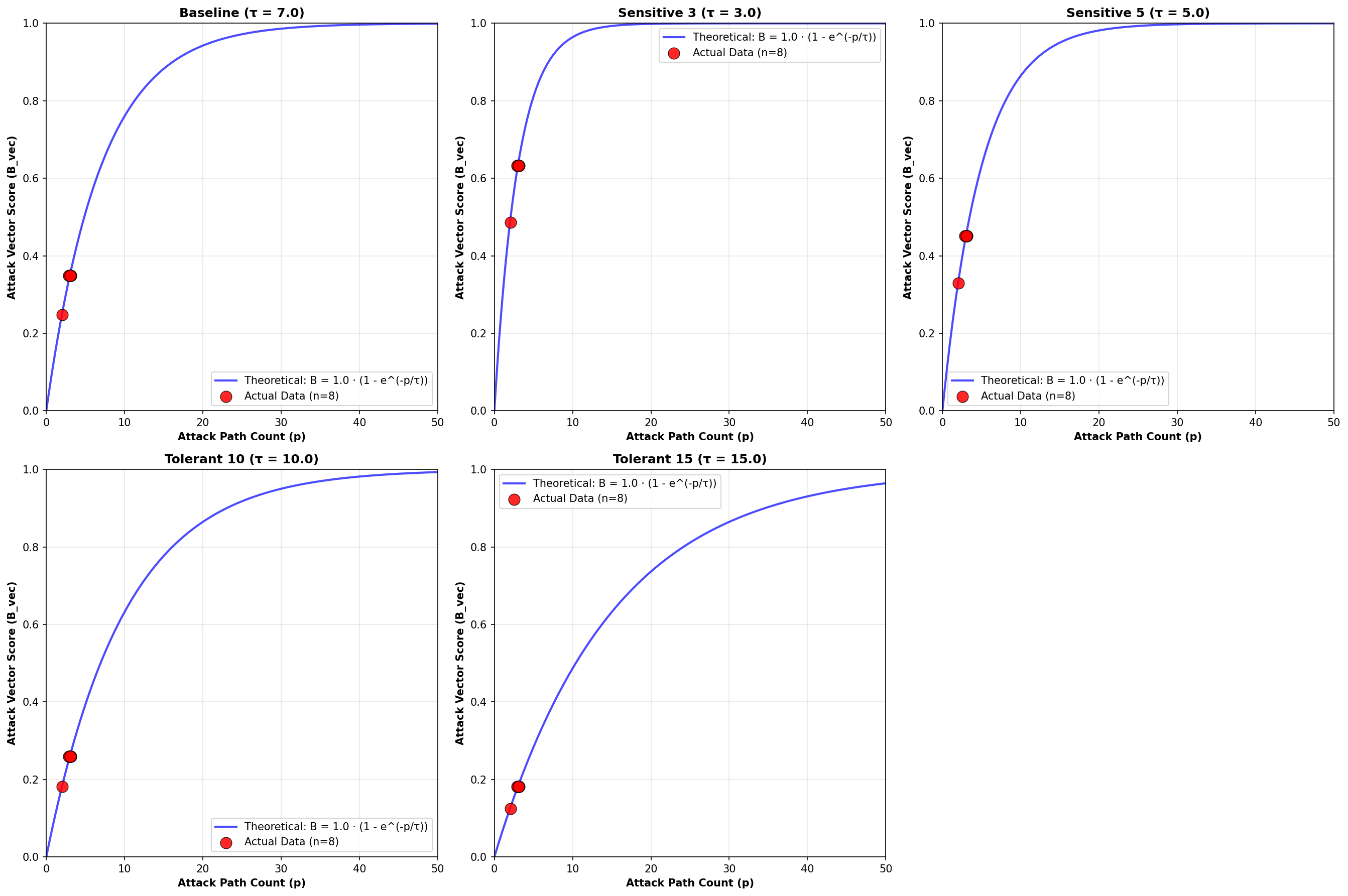}
    \caption{Sensitivity of the attack-vector exposure score to the saturation parameter $\tau$. Each panel shows the theoretical response curve, and observed resource scores.}
    \label{fig:abl3_tau_sensitivity}
\end{figure}

Figure~\ref{fig:abl3_tau_sensitivity} shows the theoretical response curve for each $\tau$ value, together with the observed resource scores. The theoretical line represents the expected score as a function of attack-path count. Its shape illustrates the intended saturation behavior: the score increases rapidly for the first few paths, but the marginal increase decreases as the path count grows. As $\tau$ increases, the curve becomes flatter, meaning that the same number of attack paths produces a lower score. Thus, the theoretical curves make the calibration role of $\tau$ explicit.

\begin{table}[t]
\centering
\small
\renewcommand{\arraystretch}{1.15}
\begin{tabular}{lcc}
\toprule
\textbf{Configuration} 
& $\boldsymbol{\tau}$
& \textbf{Resource-weighted mean $B_{\text{vec}}$} \\
\midrule
Sensitive ($\tau=3$)  & $3.0$  & $0.6139$ \\
Sensitive ($\tau=5$)  & $5.0$  & $0.4360$ \\
Baseline ($\tau=7$)    & $7.0$  & $0.3361$ \\
Tolerant ($\tau=10$)  & $10.0$ & $0.2494$ \\
Tolerant ($\tau=15$)  & $15.0$ & $0.1742$ \\
\bottomrule
\end{tabular}
\caption{Resource-weighted attack-vector scores under different values of $\tau$.}
\label{tab:tau_sensitivity}
\end{table}

Table~\ref{tab:tau_sensitivity} shows the same pattern quantitatively. The most sensitive configuration, $\tau=3$, produces the highest resource-weighted mean score $(0.6139)$, while increasingly tolerant configurations produce progressively lower scores. The mean decreases to $0.4360$ for $\tau=5$, $0.3361$ for the baseline $\tau=7$, $0.2494$ for $\tau=10$, and $0.1742$ for $\tau=15$. This confirms that $\tau$ directly controls how strongly the model reacts to attack-vector evidence.

The empirical scores follow the theoretical ordering: lower $\tau$ values consistently assign higher scores for the same attack-path counts, while higher $\tau$ values produce more tolerant scoring. Since the observed resources are concentrated in the low path-count range, most of the practical calibration occurs around the first few attack paths. This is expected because attack-vector evidence is sparse, but high-signal when present.

The baseline setting $\tau=7$ provides a balanced calibration. It is less aggressive than $\tau=3$ and $\tau=5$, which may overemphasize limited attack-vector evidence, but more responsive than $\tau=10$ and $\tau=15$, which may understate concrete exploitability. Overall, the analysis supports the attack-vector channel as a targeted exploitability signal: the theoretical curves show bounded and saturating behavior, and the empirical scores confirm that the implementation follows this behavior on resources where attack-vector evidence exists.

\subsection{Sensitivity of Contextual Modulation}
\label{sec:eval_ablation_context_modulation}

This analysis evaluates the effect of contextual modulation on the final asset score. While the previous analyses focused on the exposure channels, this analysis examines the transition from the base score $B(a)$ to the final score $s(a)$ after applying the contextual multiplier. The analysis is restricted to resources with available business-function or data-criticality context, since these are the resources for which contextual modulation is expected to have a meaningful effect.

We evaluate five contextualization-strength settings: conservative $(\alpha=0.10)$, moderate $(\alpha=0.15)$, baseline $(\alpha=0.20)$, aggressive $(\alpha=0.30)$, and very-aggressive $(\alpha=0.40)$. These settings correspond to multiplier ranges of $[0.90,1.10]$, $[0.85,1.15]$, $[0.80,1.20]$, $[0.70,1.30]$, and $[0.60,1.40]$, respectively. For each severity-weight configuration, we compare the base-score distribution with the final-score distribution across these values of $\alpha$. This allows us to observe how strongly contextual information shifts resources between prioritization ranges.

\begin{table}[ht!]
\centering
\small
\renewcommand{\arraystretch}{1.15}
\begin{tabular}{lcccc}
\toprule
\textbf{Severity-weight configuration}
& $\boldsymbol{\alpha=0.10}$
& $\boldsymbol{\alpha=0.20}$
& $\boldsymbol{\alpha=0.30}$
& $\boldsymbol{\alpha=0.40}$ \\
\midrule

Ultra-conservative $(0.0001,0.002,0.012,0.12,0.30)$
& $0.1239$ & $0.1287$ & $0.1336$ & $0.1384$ \\

Very-conservative $(0.0002,0.004,0.02,0.18,0.40)$
& $0.1586$ & $0.1644$ & $0.1703$ & $0.1761$ \\

Conservative $(0.0005,0.008,0.035,0.25,0.50)$
& $0.2023$ & $0.2093$ & $0.2164$ & $0.2235$ \\

Baseline $(0.002,0.02,0.08,0.35,0.70)$
& $0.2746$ & $0.2838$ & $0.2930$ & $0.3019$ \\

Aggressive $(0.002,0.02,0.07,0.45,0.80)$
& $0.3110$ & $0.3210$ & $0.3311$ & $0.3408$ \\

Linear $(0.005,0.05,0.15,0.40,0.70)$
& $0.3480$ & $0.3605$ & $0.3729$ & $0.3851$ \\

\bottomrule
\end{tabular}
\caption{Resource-weighted final score under increasing contextualization strength. Severity-weight tuples are ordered as $(\textsc{Info},\textsc{Low},\textsc{Medium},\textsc{High},\textsc{Critical})$.}
\label{tab:context_modulation_mean_scores}
\end{table}

Table~\ref{tab:context_modulation_mean_scores} shows that increasing $\alpha$ consistently increases the final score across all severity-weight configurations. This confirms that contextual modulation behaves monotonically: stronger contextualization increases the effect of business-function and data-criticality signals on resources where such context is available. The increase is gradual for the non-linear configurations. For example, under the baseline severity weights, the score increases from $0.2746$ at $\alpha=0.10$ to $0.2838$ at $\alpha=0.20$, $0.2930$ at $\alpha=0.30$, and $0.3019$ at $\alpha=0.40$. A similar pattern appears under the conservative, very-conservative, ultra-conservative, and aggressive mappings.

The severity-weight configuration determines the starting point from which contextual modulation operates. Under the ultra-conservative configuration, the scores are lowest across all $\alpha$ values, increasing from $0.1239$ to $0.1384$. Under the linear configuration, the scores are substantially higher, increasing from $0.3480$ to $0.3851$. This reflects the effect of the severity mapping on the base score: when the exposure channel is already elevated, contextual modulation amplifies a higher baseline.

\begin{table*}[ht!]
\centering
\scriptsize
\renewcommand{\arraystretch}{1.25}
\setlength{\tabcolsep}{4pt}
\begin{tabularx}{\textwidth}{p{2.7cm}XXXXXX}
\toprule
\textbf{Severity-weight configuration}
& \textbf{Base score}
& $\boldsymbol{\alpha=0.10}$
& $\boldsymbol{\alpha=0.15}$
& $\boldsymbol{\alpha=0.20}$
& $\boldsymbol{\alpha=0.30}$
& $\boldsymbol{\alpha=0.40}$ \\
\midrule

\makecell[l]{Ultra-conservative\\
$(0.0001,0.002,0.012,$\\$0.12,0.30)$}
& \dist{93.0}{5.0}{2.0}{0.0}{0.0}
& \dist{93.0}{5.0}{2.0}{0.0}{0.0}
& \dist{92.9}{4.9}{2.2}{0.0}{0.0}
& \dist{92.7}{5.2}{2.1}{0.0}{0.0}
& \dist{93.8}{4.6}{1.6}{0.0}{0.0}
& \dist{94.4}{3.9}{1.7}{0.0}{0.0} \\

\cmidrule(lr){1-7}

\makecell[l]{Very-conservative\\
$(0.0002,0.004,0.02,$\\$0.18,0.40)$}
& \dist{90.2}{7.1}{2.7}{0.0}{0.0}
& \dist{90.2}{7.1}{2.7}{0.0}{0.0}
& \dist{89.9}{7.3}{2.8}{0.0}{0.0}
& \dist{82.2}{15.0}{2.8}{0.0}{0.0}
& \dist{82.0}{15.2}{2.2}{0.6}{0.0}
& \dist{71.8}{25.3}{2.2}{0.1}{0.6} \\

\cmidrule(lr){1-7}

\makecell[l]{Conservative\\
$(0.0005,0.008,0.035,$\\$0.25,0.50)$}
& \dist{45.9}{50.9}{3.2}{0.0}{0.0}
& \dist{45.9}{50.9}{3.2}{0.0}{0.0}
& \dist{46.7}{49.2}{3.4}{0.6}{0.0}
& \dist{47.0}{48.8}{3.5}{0.7}{0.0}
& \dist{50.4}{45.2}{3.4}{0.3}{0.6}
& \dist{52.2}{43.4}{3.4}{0.2}{0.7} \\

\cmidrule(lr){1-7}

\makecell[l]{Baseline\\
$(0.002,0.02,0.08,$\\$0.35,0.70)$}
& \dist{35.7}{46.2}{17.4}{0.7}{0.0}
& \dist{35.7}{46.2}{17.4}{0.7}{0.0}
& \dist{35.7}{46.2}{17.2}{0.9}{0.0}
& \dist{35.7}{46.1}{17.3}{0.9}{0.0}
& \dist{35.7}{37.6}{25.6}{0.2}{0.9}
& \dist{35.7}{36.2}{26.5}{0.4}{1.1} \\

\cmidrule(lr){1-7}

\makecell[l]{Aggressive\\
$(0.002,0.02,0.07,$\\$0.45,0.80)$}
& \dist{35.7}{32.7}{30.7}{0.9}{0.0}
& \dist{35.7}{32.7}{30.7}{0.9}{0.0}
& \dist{35.7}{33.0}{30.1}{1.1}{0.0}
& \dist{35.7}{27.6}{35.5}{1.2}{0.0}
& \dist{35.7}{28.5}{34.3}{0.4}{1.1}
& \dist{35.7}{28.9}{33.7}{0.1}{1.5} \\

\cmidrule(lr){1-7}

\makecell[l]{Linear\\
$(0.005,0.05,0.15,$\\$0.40,0.70)$}
& \dist{23.7}{37.5}{38.0}{0.8}{0.0}
& \dist{23.7}{37.5}{38.0}{0.8}{0.0}
& \dist{23.8}{37.4}{37.9}{0.9}{0.0}
& \dist{13.3}{51.1}{34.3}{1.3}{0.0}
& \dist{13.3}{52.2}{32.4}{1.2}{0.9}
& \dist{13.5}{47.1}{37.0}{0.4}{2.0} \\

\bottomrule
\end{tabularx}
\caption{Resource-weighted base-score and final-score distributions for contextualized resources. Each cell reports the distribution over score bins: \textsc{I}=\textsc{Info}, \textsc{L}=\textsc{Low}, \textsc{M}=\textsc{Medium}, \textsc{H}=\textsc{High}, and \textsc{C}=\textsc{Critical}. Severity-weight tuples are ordered as $(\textsc{Info},\textsc{Low},\textsc{Medium},\textsc{High},\textsc{Critical})$.}
\label{tab:base_vs_final_contextualized}
\end{table*}

Table~\ref{tab:base_vs_final_contextualized} summarizes the resource-weighted base-score and final-score distributions for contextualized resources across severity-weight configurations and contextualization strengths. For the non-linear severity-weight configurations, the base scores are concentrated mainly in the \textsc{Info} and \textsc{Low} bins, with smaller \textsc{Medium} shares. As $\alpha$ increases, the final-score distribution shifts upward: the share of resources in the \textsc{High} and \textsc{Critical} bins increases, while the \textsc{Medium} and \textsc{Low} shares are redistributed. This indicates that contextual modulation does not uniformly inflate all scores; rather, its visible effect is concentrated in the upper tail of resources whose exposure is accompanied by strong contextual importance.

The effect is most visible at higher contextualization strengths, especially $\alpha=0.30$ and $\alpha=0.40$. Under the baseline configuration, $\alpha=0.40$ produces $1.1\%$ of resources in the \textsc{Critical} bin and $0.4\%$ in the \textsc{High} bin, accompanied by an increase in the \textsc{Medium} share from $17.4\%$ to $26.5\%$. The conservative configuration produces a similar but slightly weaker effect, with $0.7\%$ of resources reaching \textsc{Critical} at $\alpha=0.40$. The aggressive configuration produces the largest \textsc{Critical} share among the non-linear configurations, reaching $1.5\%$ at $\alpha=0.40$, because its higher severity weights generate larger base-score values that can be amplified by the multiplier. The very-conservative and ultra-conservative configurations produce weaker effects, with $0.6\%$ and $0\%$ of resources reaching \textsc{Critical} at $\alpha=0.40$, respectively, because their base-score distributions are concentrated in the \textsc{Info} bin and remain below the threshold for upper bins even after positive contextual modulation. The linear configuration produces the most pronounced upper-tail effect: at $\alpha=0.40$, $2.0\%$ of resources are assigned to the \textsc{Critical} bin, accompanied by a noticeable shift out of the \textsc{Info} bin (from $23.7\%$ to $13.5\%$). This reinforces the earlier observation that the linear severity mapping makes contextual modulation more influential because the underlying base scores are higher and more evenly spread across bins.

Overall, this analysis supports the intended behavior of contextual modulation.
The parameter $\alpha$ controls the strength of the multiplier in a predictable way: as $\alpha$ increases, the resource-weighted final score increases across all severity-weight configurations, and the upper tail of the distribution becomes more pronounced.
However, the bin-level changes are not strictly monotone for every severity bin, since resources may move from \textsc{Medium} to \textsc{High}, and from \textsc{High} to \textsc{Critical}, as the multiplier becomes stronger.
Thus, the main effect of increasing $\alpha$ is not uniform inflation, but a redistribution of contextually important resources toward higher prioritization ranges.

\section{Discussion}
\label{sec:discussion}

The proposed framework addresses a central limitation of current AI-native security systems: they provide increasingly powerful interfaces for querying security data, but they do not necessarily provide a stable basis for reasoning over that data. In modern cloud and identity environments, security posture is distributed across heterogeneous signals, including misconfigurations, permissions, attack paths, dependencies, ownership metadata, and business context. Without an intermediate abstraction that organizes these signals into a consistent representation, AI systems must reason directly over fragmented evidence. This makes prioritization sensitive to phrasing, query scope, and partial retrieval, and limits the system's ability to move from reactive assistance to proactive insight generation.

AI-native asset intelligence is proposed as such an intermediate abstraction. Its purpose is not merely to collect security data, but to transform it into a reusable intelligence layer over which both humans and AI agents can reason. The conceptual architecture separates this process into a modeling layer and a scoring layer. The modeling layer constructs a representation of assets, identities, relationships, controls, attack vectors, and blast-radius patterns. The scoring layer then converts this representation into a consistent notion of asset importance. This separation is important: reliable prioritization requires both a structured representation of the environment and a principled mechanism for comparing assets within that representation.

The scoring system operationalizes this comparison by separating intrinsic exposure from contextual importance. Intrinsic exposure captures measurable evidence that an asset is security-relevant, such as failed controls and attack-vector evidence. Contextual importance captures why that exposure matters, such as anomaly, blast radius, business-function criticality, and data criticality. This distinction prevents two common failure modes. First, it avoids treating all findings of the same severity as equally important, regardless of where they occur. Second, it prevents context from creating high priority in the absence of exposure. The final score is therefore not a direct aggregation of all available signals, but a controlled interaction between exposure and context.

The behavioral evaluation supports this design. The misconfiguration channel exhibits bounded accumulation and severity-weighted diminishing returns, allowing repeated findings to increase priority in proportion to their severity weight without causing unbounded score inflation. The attack-vector channel behaves as a targeted exploitability signal: although attack-vector evidence is sparse, its presence meaningfully increases exposure because it indicates a concrete exploitation pattern rather than a generic hygiene issue. Contextual modulation behaves as an evidence-preserving adjustment: increasing the contextualization strength shifts a controlled fraction of resources toward higher prioritization ranges, but its effect remains coupled to the base exposure induced by the severity-weight configuration. Together, these results indicate that the scoring system behaves consistently with the intended semantics of its components.

The analysis also highlights the importance of calibration. Severity weights, the attack-vector saturation parameter $\tau$, and the contextualization parameter $\alpha$ each control a distinct part of the scoring pipeline. This makes the model tunable while preserving interpretability. Different organizations may reasonably choose different operating points depending on risk tolerance, alert volume, and remediation capacity. The value of the framework is that such calibration occurs within a stable structure: changing a parameter changes a well-defined behavior, rather than altering an opaque ranking mechanism.

The role of AI-based contextualization should also be interpreted carefully. AI-based severity adjustment and semantic classification can refine prioritization, but they are most useful when embedded in a bounded scoring framework. The evaluation shows that AI-based severity adjustment primarily affects lower and middle prioritization ranges and does not broadly create extreme scores. This is desirable: AI can help interpret local context, but its effect should remain auditable and constrained by the scoring semantics. In this sense, the framework treats AI not as an unconstrained ranking oracle, but as one component in a structured reasoning pipeline.

AI-based severity adjustment is one concrete instance of this broader contextualization principle. A finding's original severity is usually defined at the control or rule level, but its operational meaning depends on the asset on which it appears. For example, public access, missing encryption, or excessive permissions may have different implications depending on whether the affected resource is a development artifact, a production data store, or part of an event-driven execution path. AI-based severity adjustment allows the system to reinterpret the finding in light of asset-specific context, rather than treating the original rule severity as fixed. This is analogous to business-function and data-criticality contextualization: in all cases, AI is used to map heterogeneous evidence into structured, bounded, and auditable scoring inputs. The important design constraint is that AI does not directly assign the final risk score. Instead, it adjusts intermediate semantic variables that are then processed by the deterministic scoring model.

The broader implication is that proactive AI for security posture management depends on stable inference. A system cannot reliably surface important assets on its own if its prioritization logic changes across prompts or partial views of the environment. By constructing a canonical asset representation and applying a consistent scoring mechanism, AI-native asset intelligence provides the conditions under which proactive behavior becomes possible. The system can continuously evaluate the environment, identify resources where exposure and context align, and surface these resources without relying on a user to formulate the exact query.

Several limitations remain. The evaluation is performed on a single production-environment snapshot from one organization. Although the snapshot is large and heterogeneous, additional environments are needed to assess generality across organizational structures, cloud architectures, and business domains. The evaluation is also behavioral rather than predictive: without incident ground truth, we cannot claim that higher-scoring resources are empirically more likely to be compromised. Finally, the contextual criteria and aggregation weights are manually specified. This improves transparency, but future work should investigate expert validation, organization-specific calibration, and learning-based adjustment of scoring parameters.

Overall, the main contribution is not only the final scalar score, but the structured reasoning process that produces it. By combining a conceptual architecture for asset intelligence with a principled scoring system, the framework provides a foundation for consistent, explainable, and proactive security-posture reasoning.

\section{Conclusion}
\label{sec:conclusion}

This paper introduced \emph{AI-native asset intelligence}, a framework for transforming fragmented security-posture data into a structured intelligence layer for asset-level reasoning. The motivation is that AI-native security assistants cannot become reliably proactive if they reason directly over disconnected findings, identities, configurations, and tool outputs. Proactive insight generation requires stable inference, and stable inference requires a canonical representation of assets and a principled basis for comparing them.

The paper makes two main contributions. First, it presents a conceptual architecture that combines a modeling layer for constructing a unified asset representation with a scoring layer for converting heterogeneous signals into asset-level priority. Second, it introduces a context-aware scoring system that separates intrinsic exposure from contextual importance. Misconfiguration findings and attack-vector evidence determine exposure, while anomaly, blast radius, business-function criticality, and data criticality determine how that exposure should be prioritized.

We evaluated the scoring system on a production-environment snapshot containing $131{,}625$ resources across 15 vendors and 178 asset types. The evaluation focused on behavioral properties through sensitivity analyses and ablations. The results show that the model responds predictably to severity-weight calibration, attack-vector saturation, AI-based severity adjustment, and contextual modulation. In particular, contextual modulation selectively elevates resources where technical exposure is reinforced by organizational importance, rather than uniformly inflating all scores.

Proactive AI for security posture management is not merely a user-interface capability, but a consequence of structured and consistent reasoning. By providing both an asset-intelligence architecture and a principled scoring mechanism, the proposed framework enables AI systems to move beyond reactive querying toward stable prioritization and proactive insight generation. Future work should evaluate the framework across additional organizations, incorporate expert validation of ranking quality, and explore adaptive calibration based on organizational risk tolerance and remediation outcomes.

\bibliographystyle{unsrt}  
\bibliography{references}

@article{liu2024misconfig,
  title={Rethinking Software Misconfigurations in the Real World: An Empirical Study and Literature Analysis},
  author={Liu, Yuhao and Zhou, Yingnan and Zhang, Hanfeng and Chang, Zhiwei and Xu, Sihan and Jia, Yan and Wang, Wei and Liu, Zheli},
  journal={arXiv preprint arXiv:2412.11121},
  year={2024}
}

@article{shao2026framework,
  title={Design and Implementation of an Open-Source Security Framework for Cloud Infrastructure},
  author={Shao, Wanru},
  journal={arXiv preprint arXiv:2604.03331},
  year={2026}
}

@misc{first_cvss40,
  author       = {{Forum of Incident Response and Security Teams}},
  title        = {Common Vulnerability Scoring System version 4.0: Specification Document},
  year         = {2023},
  note         = {Document Version 1.2},
  url          = {https://www.first.org/cvss/specification-document}
}

@misc{first_epss_2026,
  author       = {{Forum of Incident Response and Security Teams}},
  title        = {Exploit Prediction Scoring System (EPSS)},
  year         = {2026},
  note         = {Accessed 2026-04-26},
  url          = {https://www.first.org/epss/}
}

@article{jacobs_epss_2021,
  author       = {Jacobs, Jay and Romanosky, Sasha and Edwards, Benjamin and Roytman, Michael and Adjerid, Idris},
  title        = {Exploit Prediction Scoring System (EPSS)},
  journal      = {Digital Threats: Research and Practice},
  volume       = {2},
  number       = {3},
  pages        = {1--17},
  year         = {2021},
  doi          = {10.1145/3436242},
  url          = {https://doi.org/10.1145/3436242}
}

@article{howland_cvss_2023,
  author       = {Howland, Henry},
  title        = {CVSS: Ubiquitous and Broken},
  journal      = {Digital Threats: Research and Practice},
  volume       = {4},
  number       = {1},
  pages        = {1:1--1:12},
  year         = {2023},
  doi          = {10.1145/3491263},
  url          = {https://doi.org/10.1145/3491263}
}

@techreport{spring_ssvc_2021,
  author       = {Spring, Jonathan and Householder, Allen D. and Hatleback, Eric and Manion, Art and Oliver, Madison and Sarvepalli, Vijay S. and Tyzenhaus, Laurie and Yarbrough, Charles G.},
  title        = {Prioritizing Vulnerability Response: A Stakeholder-Specific Vulnerability Categorization (Version 2.0)},
  institution  = {Software Engineering Institute, Carnegie Mellon University},
  year         = {2021},
  month        = apr,
  url          = {https://www.sei.cmu.edu/library/prioritizing-vulnerability-response-a-stakeholder-specific-vulnerability-categorization-version-20/}
}

@techreport{paulsen_criticality_2018,
  author       = {Paulsen, Celia and Boyens, Jon and Bartol, Nadya and Winkler, Kris},
  title        = {Criticality Analysis Process Model: Prioritizing Systems and Components},
  institution  = {National Institute of Standards and Technology},
  number       = {NIST IR 8179},
  year         = {2018},
  doi          = {10.6028/NIST.IR.8179},
  url          = {https://doi.org/10.6028/NIST.IR.8179}
}

@techreport{quinn_8286b_2025,
  author       = {Quinn, Stephen D. and Ivy, Nahla and Barrett, Matthew and Witte, Gregory A. and Gardner, Robert K.},
  title        = {Prioritizing Cybersecurity Risk for Enterprise Risk Management},
  institution  = {National Institute of Standards and Technology},
  number       = {NIST IR 8286B-upd1},
  year         = {2025},
  doi          = {10.6028/NIST.IR.8286B-upd1},
  url          = {https://doi.org/10.6028/NIST.IR.8286B-upd1}
}

@techreport{quinn_8286d_2025,
  author       = {Quinn, Stephen D. and Ivy, Nahla and Chua, Julie and Barrett, Matthew and Feldman, Larry and Topper, Daniel and Witte, Gregory A. and Gardner, Robert K.},
  title        = {Using Business Impact Analysis to Inform Risk Prioritization and Response},
  institution  = {National Institute of Standards and Technology},
  number       = {NIST IR 8286D-upd1},
  year         = {2025},
  doi          = {10.6028/NIST.IR.8286D-upd1},
  url          = {https://doi.org/10.6028/NIST.IR.8286D-upd1}
}

@article{kure_assetcriticality_2022,
  author       = {Kure, Halima Ibrahim and Islam, Shareeful and Ghazanfar, M. and Raza, Ali and Pasha, Mohsin},
  title        = {Asset criticality and risk prediction for an effective cybersecurity risk management of cyber-physical system},
  journal      = {Neural Computing and Applications},
  volume       = {34},
  number       = {1},
  pages        = {493--514},
  year         = {2022},
  doi          = {10.1007/s00521-021-06400-0},
  url          = {https://doi.org/10.1007/s00521-021-06400-0}
}

@article{elmiger_graphs_2024,
  author       = {Elmiger, Marius and Lemoudden, Mouad and Pitropakis, Nikolaos and Buchanan, William J.},
  title        = {Start thinking in graphs: using graphs to address critical attack paths in a Microsoft cloud tenant},
  journal      = {International Journal of Information Security},
  volume       = {23},
  number       = {1},
  pages        = {467--485},
  year         = {2024},
  doi          = {10.1007/s10207-023-00751-6},
  url          = {https://doi.org/10.1007/s10207-023-00751-6}
}

@inproceedings{polinsky_grasp_2024,
  author       = {Polinsky, Isaac and Datta, Pubali and Bates, Adam and Enck, William},
  title        = {GRASP: Hardening Serverless Applications through Graph Reachability Analysis of Security Policies},
  booktitle    = {Proceedings of the ACM Web Conference 2024},
  pages        = {1644--1655},
  year         = {2024},
  doi          = {10.1145/3589334.3645436},
  url          = {https://doi.org/10.1145/3589334.3645436}
}

@article{gouglidis_googleiam_2023,
  author       = {Gouglidis, Antonios and Kagia, Anna and Hu, Vincent C.},
  title        = {Model Checking Access Control Policies: A Case Study using Google Cloud IAM},
  journal      = {CoRR},
  volume       = {abs/2303.16688},
  year         = {2023},
  url          = {https://arxiv.org/abs/2303.16688}
}

@article{glockler_iam_2024,
  author       = {Gl{\"o}ckler, Jana and Sedlmeir, Johannes and Frank, Muriel and Fridgen, Gilbert},
  title        = {A Systematic Review of Identity and Access Management Requirements in Enterprises and Potential Contributions of Self-Sovereign Identity},
  journal      = {Business \& Information Systems Engineering},
  volume       = {66},
  number       = {4},
  pages        = {421--440},
  year         = {2024},
  doi          = {10.1007/s12599-023-00830-x},
  url          = {https://doi.org/10.1007/s12599-023-00830-x}
}

@article{yan_graphsurvey_2024,
  author       = {Yan, Bo and Yang, Cheng and Shi, Chuan and Fang, Yong and Li, Qi and Ye, Yanfang and Du, Junping},
  title        = {Graph Mining for Cybersecurity: A Survey},
  journal      = {ACM Computing Surveys},
  year         = {2024},
  doi          = {10.1145/3610228},
  url          = {https://doi.org/10.1145/3610228}
}

@article{jung_cavp_2022,
  author       = {Jung, Bill and Li, Yan and Bechor, Tamir},
  title        = {CAVP: A context-aware vulnerability prioritization model},
  journal      = {Computers \& Security},
  volume       = {116},
  pages        = {102639},
  year         = {2022},
  doi          = {10.1016/j.cose.2022.102639},
  url          = {https://doi.org/10.1016/j.cose.2022.102639}
}

@article{yadav_smartpatch_2022,
  author       = {Yadav, Geeta and Gauravaram, Praveen and Jindal, Arun Kumar and Paul, Kolin},
  title        = {SmartPatch: A patch prioritization framework},
  journal      = {Computers in Industry},
  volume       = {137},
  pages        = {103595},
  year         = {2022},
  doi          = {10.1016/j.compind.2021.103595},
  url          = {https://doi.org/10.1016/j.compind.2021.103595}
}

@article{jiang_vulnprior_2025,
  author       = {Jiang, Yuning and Oo, Nay and Meng, Qiaoran and Lim, Hoon Wei and Sikdar, Biplab},
  title        = {A Survey on Vulnerability Prioritization: Taxonomy, Metrics, and Research Challenges},
  journal      = {CoRR},
  volume       = {abs/2502.11070},
  year         = {2025},
  doi          = {10.48550/arXiv.2502.11070},
  url          = {https://arxiv.org/abs/2502.11070}
}

@article{sikos_cyberkg_2023,
  author       = {Sikos, Leslie F.},
  title        = {Cybersecurity knowledge graphs},
  journal      = {Knowledge and Information Systems},
  volume       = {65},
  pages        = {3511--3531},
  year         = {2023},
  doi          = {10.1007/s10115-023-01860-3},
  url          = {https://doi.org/10.1007/s10115-023-01860-3}
}

@article{zhao_ckg_2024,
  author       = {Zhao, Xiaojuan and Jiang, Rong and Han, Yue and Li, Aiping and Peng, Zhichao},
  title        = {A survey on cybersecurity knowledge graph construction},
  journal      = {Computers \& Security},
  volume       = {136},
  pages        = {103524},
  year         = {2024},
  doi          = {10.1016/j.cose.2023.103524},
  url          = {https://doi.org/10.1016/j.cose.2023.103524}
}

@inproceedings{falcarin_cybergraph_2024,
  author       = {Falcarin, Paolo and Dainese, Fabio},
  title        = {Building a Cybersecurity Knowledge Graph with CyberGraph},
  booktitle    = {Proceedings of the 2024 ACM/IEEE 4th International Workshop on Engineering and Cybersecurity of Critical Systems and the 2024 IEEE/ACM Second International Workshop on Software Vulnerability},
  year         = {2024},
  doi          = {10.1145/3643662.3643962},
  url          = {https://doi.org/10.1145/3643662.3643962}
}

@article{cheng_ctinexus_2024,
  author       = {Cheng, Yutong and Bajaber, Osama and Tsegai, Saimon Amanuel and Song, Dawn and Gao, Peng},
  title        = {CTINEXUS: Leveraging Optimized LLM In-Context Learning for Constructing Cybersecurity Knowledge Graphs Under Data Scarcity},
  journal      = {CoRR},
  volume       = {abs/2410.21060},
  year         = {2024},
  doi          = {10.48550/arXiv.2410.21060},
  url          = {https://arxiv.org/abs/2410.21060}
}

@inproceedings{luo_rog_2024,
  author       = {Luo, Linhao and Li, Yuan-Fang and Haffari, Gholamreza and Pan, Shirui},
  title        = {Reasoning on Graphs: Faithful and Interpretable Large Language Model Reasoning},
  booktitle    = {The Twelfth International Conference on Learning Representations},
  year         = {2024},
  url          = {https://proceedings.iclr.cc/paper_files/paper/2024/hash/3e2aeb66481dd63a32421bf032b70384-Abstract-Conference.html}
}

@inproceedings{ji_kgllm_2024,
  author       = {Ji, Yixin and Wu, Kaixin and Li, Juntao and Chen, Wei and Zhong, Mingjie and Xu, Jia and Zhang, Min},
  title        = {Retrieval and Reasoning on KGs: Integrate Knowledge Graphs into Large Language Models for Complex Question Answering},
  booktitle    = {Findings of the Association for Computational Linguistics: EMNLP 2024},
  pages        = {7598--7610},
  year         = {2024},
  url          = {https://aclanthology.org/2024.findings-emnlp.446/}
}

@article{cheng_neuralsymbolic_2024,
  author       = {Cheng, Kewei and Ahmed, Nesreen K. and Rossi, Ryan A. and Willke, Theodore L. and Sun, Yizhou},
  title        = {Neural-Symbolic Methods for Knowledge Graph Reasoning: A Survey},
  journal      = {ACM Transactions on Knowledge Discovery from Data},
  volume       = {18},
  number       = {9},
  pages        = {225:1--225:44},
  year         = {2024},
  doi          = {10.1145/3686806},
  url          = {https://doi.org/10.1145/3686806}
}

@article{xu_llm4security_2025,
  author       = {Xu, Hanxiang and Wang, Shenao and Li, Ningke and Wang, Kailong and Zhao, Yanjie and Chen, Kai and Yu, Ting and Liu, Yang and Wang, Haoyu},
  title        = {Large Language Models for Cyber Security: A Systematic Literature Review},
  journal      = {ACM Computing Surveys},
  year         = {2025},
  doi          = {10.1145/3769676},
  url          = {https://doi.org/10.1145/3769676}
}

@inproceedings{banstola_agentic_2026,
  author       = {Banstola, Kritan and Al Faisal, Faayed and Ou, Xinming},
  title        = {Experiences of Using Agentic AI to Fill Tooling Gaps in a Security Operations Center},
  booktitle    = {Workshop on SOC Operations and Construction (WOSOC), co-located with the NDSS Symposium},
  year         = {2026},
  url          = {https://www.ndss-symposium.org/ndss-paper/auto-draft-735/}
}

@article{wei_cortex_2025,
  author       = {Wei, Bowen and Tay, Yuan Shen and Liu, Howard and Pan, Jinhao and Luo, Kun and Zhu, Ziwei and Jordan, Chris},
  title        = {CORTEX: Collaborative LLM Agents for High-Stakes Alert Triage},
  journal      = {CoRR},
  volume       = {abs/2510.00311},
  year         = {2025},
  doi          = {10.48550/arXiv.2510.00311},
  url          = {https://arxiv.org/abs/2510.00311}
}

@article{zhou_llmpd_2024,
  author       = {Zhou, Yuyang and Cheng, Guang and Du, Kang and Chen, Zihan},
  title        = {Toward Intelligent and Secure Cloud: Large Language Model Empowered Proactive Defense},
  journal      = {CoRR},
  volume       = {abs/2412.21051},
  year         = {2024},
  doi          = {10.48550/arXiv.2412.21051},
  url          = {https://arxiv.org/abs/2412.21051}
}

@article{ali_securecai_2026,
  author       = {Ali, Mohammed Himayath and Abdullah, Mohammed Aqib and Uddin, Mohammed Mudassir and Alam, Shahnawaz},
  title        = {SecureCAI: Injection-Resilient LLM Assistants for Cybersecurity Operations},
  journal      = {CoRR},
  volume       = {abs/2601.07835},
  year         = {2026},
  doi          = {10.48550/arXiv.2601.07835},
  url          = {https://arxiv.org/abs/2601.07835}
}

@inproceedings{zhuo_prosa_2024,
  author       = {Zhuo, Jingming and Zhang, Songyang and Fang, Xinyu and Duan, Haodong and Lin, Dahua and Chen, Kai},
  title        = {ProSA: Assessing and Understanding the Prompt Sensitivity of LLMs},
  booktitle    = {Findings of the Association for Computational Linguistics: EMNLP 2024},
  pages        = {1950--1976},
  year         = {2024},
  url          = {https://aclanthology.org/2024.findings-emnlp.108/}
}

@inproceedings{polo_prompteval_2024,
  author       = {Polo, Felipe Maia and Xu, Ronald and Weber, Lucas and Silva, M{\'i}rian and Bhardwaj, Onkar and Choshen, Leshem and de Oliveira, Allysson Flavio Melo and Sun, Yuekai and Yurochkin, Mikhail},
  title        = {Efficient multi-prompt evaluation of LLMs},
  booktitle    = {Advances in Neural Information Processing Systems 37},
  year         = {2024},
  url          = {https://openreview.net/forum?id=jzkpwcj200}
}

@inproceedings{hua_promptartifact_2025,
  author       = {Hua, Andong and Tang, Kenan and Gu, Chenhe and Gu, Jindong and Wong, Eric and Qin, Yao},
  title        = {Flaw or Artifact? Rethinking Prompt Sensitivity in Evaluating LLMs},
  booktitle    = {Proceedings of the 2025 Conference on Empirical Methods in Natural Language Processing},
  pages        = {19889--19899},
  year         = {2025},
  doi          = {10.18653/v1/2025.emnlp-main.1006},
  url          = {https://aclanthology.org/2025.emnlp-main.1006/}
}

@misc{mitre_attack,
  title        = {{MITRE ATT\&CK}: Adversarial Tactics, Techniques, and Common Knowledge},
  author       = {{MITRE Corporation}},
  year         = {2024},
  howpublished = {\url{https://attack.mitre.org/}},
  note         = {Accessed: 2026-04-26}
}

@misc{mitre_cwe,
  title        = {{Common Weakness Enumeration (CWE)}},
  author       = {{MITRE Corporation}},
  year         = {2024},
  howpublished = {\url{https://cwe.mitre.org/}},
  note         = {Accessed: 2026-04-26}
}

@misc{mitre_capec,
  title        = {{Common Attack Pattern Enumeration and Classification (CAPEC)}},
  author       = {{MITRE Corporation}},
  year         = {2024},
  howpublished = {\url{https://capec.mitre.org/}},
  note         = {Accessed: 2026-04-26}
}

@misc{engelberg2026solaVisibilityISPM,
  title        = {Sola-Visibility-ISPM: Benchmarking Agentic AI for Identity Security Posture Management Visibility},
  author       = {Engelberg, Gal and Koutsyi, Konstantin and Goldberg, Leon and Elezra, Reuven and Pinto, Idan and Moalem, Tal and Cohen, Shmuel and Weintrob, Yoni},
  year         = {2026},
  eprint       = {2601.07880},
  archivePrefix= {arXiv},
  primaryClass = {cs.CR},
  doi          = {10.48550/arXiv.2601.07880}
}

\appendix

\end{document}